\documentclass[11pt]{article}
\usepackage[fleqn]{amsmath}
\usepackage{amsthm,amssymb}
\usepackage{graphicx}
\usepackage{hhline}
\usepackage{cite}

\makeatletter
\@addtoreset{equation}{section}
\makeatother

{\theoremstyle{remark}
}

\def\rd{ {\rm d}}
\def\re{ {\rm e}}

\def\ri{{\rm i}}

\def\re{\mathrm{e}}
\def\ri{\mathrm{i}}

\def\be{\begin{equation}}
\def\ee{\end{equation}}
\def\bal{\begin{align}}
\def\eal{\end{align}}
\def\bea{\begin{eqnarray}}
\def\eea{\end{eqnarray}}

\usepackage{euscript}

\def\zb{\bar z}
\def\pz{\partial_z}
\def\pzb{\partial_{\bar z}}
\def\wb{\bar w}

\begin{document}


\begin{titlepage}

\begin{flushright}
RU-NHETC-2010-10\\
\end{flushright}

\vspace{1.5cm}

\begin{center}
\begin{LARGE}
{\bf Quantum Sine(h)-Gordon Model\\
\vspace{0.2cm}

 and \\
\vspace{0.2cm}
 Classical Integrable Equations}

\end{LARGE}

\vspace{1.3cm}

\begin{large}

{\bf S.~L.~Lukyanov and
A.~B.~Zamolodchikov}

\end{large}

\vspace{1.cm}

NHETC, Department of Physics and Astronomy\\
     Rutgers University\\
     Piscataway, NJ 08855-0849, USA\\

\vspace{.3cm}
and
\vspace{.3cm}

L.D. Landau Institute for Theoretical Physics\\
  Chernogolovka, 142432, Russia

\end{center}

\vspace{1.cm}

\begin{center}
\centerline{\bf Abstract} \vspace{.8cm}
\parbox{15.5cm}
{ We study a family of classical solutions of modified sinh-Gordon equation,
$\partial_z\partial_{{\bar z}} \eta-\re^{2\eta}+p(z)\,p({\bar z})\ \re^{-2\eta}=0$ with
$p(z)=z^{2\alpha}-s^{2\alpha}$. We show that certain connection coefficients
for solutions of the associated linear problem coincide with
the $Q$-function of the quantum sine-Gordon $(\alpha>0)$ or
sinh-Gordon $(\alpha<-1)$ models.}
\end{center}

\vspace{0.1cm}

\begin{flushleft}
\rule{3.1 in}{.01 in}\\
{March  2010}
\end{flushleft}
\vfill
\end{titlepage}
\newpage

\tableofcontents

\newpage

\begin{flushright}
{\it To the memory of Alyosha Zamolodchikov}\\
\end{flushright}

\section{ Introduction}

In over three decades of study of quantum integrable systems, a
remarkable (and largely mysterious) relation to classical
integrable equations was observed in a number of different
contexts. The first such relation was discovered long ago by
Barouch, Tracy, and McCoy\cite{Wu:1975mw},
who have derived the
spin-spin correlation function in the Ising Field Theory in terms
of special Painlev\'e III transcendent, i.e. the solution of the
differential equation,
\begin{eqnarray}\label{painleve}
\frac{\rd^2 f}{\rd \tau^2}+\frac{1}{\tau}\frac{\rd f}{\rd \tau}-\frac{1}{2}\ \sinh(2f) =0
\end{eqnarray}
which decays at $\tau\to\infty$ and is singular as $-\log(-\tau\log\tau)$
at $\tau\to 0$. The derivation relies on the free-field
nature of the Ising Field Theory, and it is still unknown if this
powerful result can be generalized in any simple way to
interacting integrable quantum field theories.

However, relation to classical integrable equation has surfaced
later in analysis of the dilute self-avoiding polymer problem
\cite{Fendley:1992jy, Zamolodchikov:1994uw},
which is related to quantum sine-Gordon
model at special value of the coupling parameter. It was found
that the off-critical partition function of a self-avoiding
polymer loop on an infinite cylinder is expressed exactly through
another solution of the same Painlev\'e III equation
\eqref{painleve}, this time with the singularity
$-\frac{1}{3}\log t +O(1)$ at small $t$. This elegant result can
be attributed to supersymmetry of the problem -- it can
be reformulated in terms of quantum sine-Gordon model at special
value of its  coupling constant  ($\beta^2=\frac{2}{3}$ in \eqref{sg}
below), where it exhibits ${\cal N}=2$ supersymmetry. Indeed, the
derivation in \cite{Cecotti:1992qh}  has its roots in deep analysis of 2D ${\cal N}=2$
supersymmetric field theories \cite{Cecotti:1991me}. At the same time, the
finite-size partition function is generally much simpler object
as compared to the correlation functions. In particular,
TBA  technique \cite{Yang:1968rm} can be employed to
determine this quantity in any integrable quantum field theory
(supersymmetric or not) as long as its S-matrix is
known \cite{Zamolodchikov:1991et}.
Therefore one may expect that some more or less
direct extension of the results of \cite{Fendley:1992jy, Zamolodchikov:1994uw} to
generic integrable theories is possible.

In this paper we propose such extension to the case of the quantum
sine-Gordon model
\begin{eqnarray}\label{sg}
\mathcal{L}=\frac{1}{16\pi}\ \left[\, \left(\partial_t\varphi\right)^2-
\left(\partial_x\varphi\right)^2\, \right] +
2\mu\,\cos\left(\beta\varphi\right)
\end{eqnarray}
at generic values of the coupling parameter $\beta^2<1$. Here $\mu$
sets the mass scale, $\mu \sim \text{[\,mass\,]}^{2-2\beta^2}$. We
will consider the theory in finite-size geometry, with the
spatial coordinate $x$ in $\varphi=\varphi(x,t)$ compactified on
a circle of a circumference $R$, with the periodic boundary
conditions
\begin{eqnarray}\label{pbc}
\varphi(x+R,t)=\varphi(x,t)\,.
\end{eqnarray}
Due to the periodicity of the potential term $2\mu\,\cos(\beta\varphi)$ in \eqref{sg} in
$\varphi$, the
space of states $\mathcal{H}$ splits into orthogonal subspaces
$\mathcal{H}_k$, characterized by the ``quasi-momentum'' $k$,
\begin{eqnarray}\label{quasi}
\varphi \to \varphi +2\pi/\beta \,: \qquad \mid \Psi_k\,\rangle \
\to\ \re^{2\pi \ri\,k}\,\mid \Psi_k \,\rangle
\end{eqnarray}
for $\mid \Psi_k\,\rangle \in \mathcal{H}_k$. We call $k$-vacuum
the ground-state of the finite-size system \eqref{sg} in the
sector $\mathcal{H}_k$.

The quantum field theory \eqref{sg} is integrable, in particular
it has infinite set of commuting local Integrals of Motion (IM)
$\mathbb{I}_{2n-1},\ \mathbb{\bar I}_{2n-1}$, $2n=2,\, 4,\,6,\,\ldots$
being the Lorentz spins of the associated local densities \cite{Kulish:1975ba}.
Of primary interest are the $k$-vacuum
eigenvalues $I_{2n-1}(k\,|\,R),\ {\bar I}_{2n-1}(k\,|\,R)$, especially
the $k$-vacuum energy $I_1 +{\bar I}_1$. In principle, these
quantities are accessible through the Thermodynamic Bethe Ansatz (TBA)  technique, but the most
efficient approach is the Destri-De Vega (DDV) equation
\cite{Destri:1992qk, Destri:1997yz} (Similar equation was earlier derived in 
the lattice $XXZ$-model in 
Ref.\cite{Klumper:1991}).
The later determines the so-called $Q$-function $Q(\theta, k\,|\,R)$,
whose asymptotic expansions at $\theta\to + \infty$ and
$\theta\to -\infty$ generate the eigenvalues $I_{2n-1}$ and
${\bar I}_{2n-1}$ (along with the eigenvalues of the nonlocal integrals
of motion of Ref.\cite{Bernard:1990ys}), respectively. Remarkable observation of
\cite{Fendley:1992jy,Zamolodchikov:1994uw} is that
at the special value $\beta^2=\frac{2}{ 3}$
these essentially quantum characteristics can be related to the
solution of the classical nonlinear differential equation
\eqref{painleve}.

Equation \eqref{painleve} of course is the radial equation
for rotationally symmetric solutions of 2D sinh-Gordon equation.
We will argue that for generic $\beta^2$ similar relation exists
to the classical Modified Sinh-Gordon equation (MShG)
\begin{eqnarray}\label{shgz}
\pz\pzb\eta -\re^{2\eta}+p(z)\,{p}(\zb)\ \re^{-2\eta}=0
\end{eqnarray}
with the functions $p(z)$ of the form
\begin{eqnarray}\label{kskssls}
p(z) = z^{2\alpha}-s^{2\alpha}\,.
\end{eqnarray}
Here $\alpha$ and $s$ are real, positive parameters\footnote{In fact, for
technical reasons in this work we limit our attention to the case
$\alpha\geq 1$, which corresponds to $\beta^2\leq\frac{1}{2}$ in\ \eqref{sg}. However,
our main results remain valid at any positive $\alpha$.}, related to the
parameters $\beta$, $\mu$ in \eqref{sg} as follows
\begin{eqnarray}
\alpha = \beta^{-2}-1\,,
\qquad\qquad  s =\Big(\frac{R}{ \pi\beta^2}\Big)^{\beta^2}\ \bigg[\,
\frac{\mu\pi\Gamma(1-\beta^2)}{ \Gamma(\beta^2)}\,\bigg]^{\frac{\beta^2}{ 2-2\beta^2}}\ ,
\end{eqnarray}
but $z,\ \zb$ are formal variables, not related to the space-time coordinates $(x,\,t)$
in \eqref{sg}.
Equation \eqref{shgz} with general $p(z)$ is well known in differential geometry (see \cite{Babenko} for
review). The same equation with polynomial $p(z)$ has emerged lately in
different contexts in SUSY gauge theories
\cite{Gaiotto:2008cd,Gaiotto:2009hg,Alday:2009dv,Alday:2009yn,Alday:2010vh},
and these later papers have  inspired to large extent  the
work presented here.

Obviously, MShG equation in general has no rotational symmetry.
Instead, it has discrete
symmetry
\begin{eqnarray}
z \to \text{e}^\frac{\ri\pi}{\alpha}\ z
\,, \qquad \zb \to \text{e}^{-\frac{\ri\pi}{\alpha}}\ {\bar z}\  .
\end{eqnarray}
We will consider solutions of MShG equation \eqref{shgz} which
respect this symmetry, continuous at all finite nonzero $z,\zb$, and
grow slower then exponential at $z,\zb\to\infty$ (precise
conditions are listed in  Section 2).
There is one-parameter set of such solutions, characterized
by the behavior $\eta \to l\,\log(z\zb)$ at $z,\zb\to 0$, with
real $l\in [-\frac{1}{2},\,\frac{1}{2}\, ]$ which will turn out to be related to the
quasi-momentum as
\bea\label{llsslasa}
l=2\,|k|-{\textstyle\frac{1}{2}}\ .
\eea

As is well known, the MShG equation is integrable, and the
associated linear problem has the form \cite{Faddeev:1987ph}
\begin{eqnarray}\label{lp}
{\boldsymbol D}{\bf\Psi}=0\,, \qquad {\bar {\boldsymbol D}}{\bf\Psi}=0\ ,
\end{eqnarray}
where ${\boldsymbol D},\, {\bar {\boldsymbol D}}$ are components of  $sl(2)$ connection\footnote{
$\sigma^a$ are the usual Pauli matrices, i.e.,
$\sigma^3=
\begin{pmatrix}
1&0\\
0&-1
\end{pmatrix},\   \sigma^+=
\begin{pmatrix}
0&1\\
0&0
\end{pmatrix},\
\sigma^-=
\begin{pmatrix}
0&0\\
1&0
\end{pmatrix}$\ .}
\begin{eqnarray}\label{ssoiso}
{\boldsymbol D}&=&\partial_z +
{\textstyle\frac{1}{ 2}}\ \partial_z\eta\ \sigma^3-\re^{\theta}\ \  \big[\,
\sigma^+\ \re^{\eta}+ \sigma^-\ p(z)\,\re^{-\eta}\, \big]\ ,\\
{\bar {\boldsymbol D}} &=&
{\partial}_{\zb} -
{\textstyle\frac{1}{2}}\ { \partial}_{\zb}\eta\ \sigma^3-
 \re^{-\theta}\, \big[\,
\sigma^-\ \re^{\eta}+ \sigma^+\ p({\bar z})\,\re^{-\eta}\, \big]\ ,
 \nonumber
\end{eqnarray}
with the spectral parameter $\theta$. The $Q$-function of the
quantum sine-Gordon model \eqref{sg} will be related to connection
coefficients for certain solution of the linear problem
\eqref{lp}.
Therefore our result can be regarded as generalization to $\mu\not=0$
of the relation \cite{Dorey:1998pt} between
the integrable structures of CFT
\cite{Bazhanov:1994ft,Bazhanov:1996dr,Bazhanov:1998dq} and spectral characteristics of
linear ordinary differential equations \cite{Voros:1992,Voros:1999bz}.
Indeed, our derivation follows very closely the
analysis in \cite{Bazhanov:1998wj}. The novel feature of the massive case is
that the coefficients in the linear problem are not elementary
functions but rather solutions of integrable nonlinear partial
differential equation \eqref{shgz}.

Similar relation exists between certain solution of the MShG equation
\eqref{shgz},
with $p(z)$ of the same form \eqref{kskssls} but this time with $\alpha < -1$,
\bea\label{alphab}
\alpha = -b^{-2}-1\ , \qquad
s =\Big(\frac{R}{ \pi b^2}\Big)^{-b^2}\ \bigg[-
\frac{ \mu\pi\Gamma(1+b^2)}{ \Gamma(-b^2)}\,\bigg]^{\frac{b^2}{ 2+2 b^2}}
\eea
and the vacuum $Q$-function of the quantum sinh-Gordon model
\begin{eqnarray}\label{shg}
\mathcal{L}=\frac{1}{16\pi}\ \left[\, \left(\partial_t\varphi\right)^2-
\left(\partial_x\varphi\right)^2\, \right] -
2\mu\,\cosh\left(b\varphi\right)\ ,
\end{eqnarray}
where again the spatial finite size geometry \eqref{pbc} is assumed. Of course, physical content
of the
quantum sinh-Gordon model
is much different from the sine-Gordon model
\eqref{sg}. In particular, \eqref{shg}
has unique vacuum. Correspondingly,
the MShG equation \eqref{shgz} with $\alpha<-1$ has unique solution
which is continuous at all finite nonzero $z, \zb$. The vacuum $Q$-function of \eqref{shg}
\cite{Zamolodchikov:2000kt, Lukyanov:2000jp,Teschner:2007ng} will be related
to the linear problem \eqref{lp} associated with this unique solution.

The paper is organized as follows.
In Section\,\ref{LInP} we discuss the MShG equation with $\alpha>0$.
We define a family
of regular solutions,
and describe their basic properties. We also discuss the associated linear problem \eqref{lp},
and define the functions $Q_{+}(\theta),\ Q_{-}(\theta)$ as certain connection coefficients.
In Section\,\ref{Sect2} we describe
how the function $Q(\theta,k)$ is constructed out
of these coefficients, and list its basic properties. In
particular, we show that it is
determined by unique solution of complex nonlinear integral equation identical
to DDV equation, and thus
coincides with the vacuum $Q$-function of the sine-Gordon model.
We also establish
relation between the classical local IM of MShG, and vacuum eigenvalues of the quantum
IM of the sine-Gordon model.
In Section\,\ref{Tfun} we define the functions $T_j(\theta,k)$ in terms of the
monodromy  of the linear
problem \eqref{lp}, and show that they coincide with the vacuum $T$-functions of the model \eqref{sg}.
In particular, at integer values of $2\alpha$ 
these functions can be determined through the finite system
of
TBA equations, which were previously derived in similar context in
\cite{Gaiotto:2008cd,Gaiotto:2009hg,Alday:2009dv,Alday:2009yn,Alday:2010vh}.
The relation is similar to that described in Ref.\cite{Dorey:1999uk} in the massless case.
Section\,\ref{GLM} is devoted 
to analysis of the inverse scattering problem in \eqref{lp}. In particular, we
present explicit series-like representation of the solution $\eta$. MShG with $\alpha<-1$, and its relation
to the quantum sinh-Gordon model \eqref{shg}, is discussed in Section\,\ref{MShSh}.

\section{\label{LInP} MShG equation and linear problem}

Although generally  $z$ and $\zb$ in \eqref{shgz} can
be regarded as independent complex variables, in the present
discussion we usually (but not always) assume them to be complex coordinates on 2D real
space. Thus, $\eta$ in \eqref{shgz} is assumed to be a function of
two real variables, $\eta=\eta(\rho,\phi)$, the polar coordinates
associated with $(z,\zb)$,
\begin{eqnarray}
z = \rho\,\re^{\ri\phi}\,,\qquad \qquad \zb = \rho\,\re^{-\ri\phi}\, .
\end{eqnarray}
We consider special family of solutions of \eqref{shgz},
parameterized by real $l \in \big[-\frac{1}{2},\frac{1}{2}\big]$, defined
by the following properties:

{\bf i}) Periodicity
\begin{eqnarray}
\eta\big(\rho, \phi+{\textstyle\frac{\pi}{\alpha}}\big) = \eta(\rho,\phi)\,.
\end{eqnarray}
In other words, the solutions $\eta(\rho,\phi)$ are single-valued functions
on a cone with the apex angle $\frac{\pi}{\alpha}$,
\begin{eqnarray}\label{cone}
{\mathbb C}_{\frac{\pi}{\alpha}}\,: \quad \phi \sim \phi+{\textstyle \frac{\pi}{\alpha}}\,, \qquad
0 \leq  \rho < \infty\,.
\end{eqnarray}

{\bf ii}) $\eta(\rho,\phi)$ are real-valued and finite everywhere on
the cone $\mathbb{C}_\frac{\pi}{\alpha}$,
except for the apex
$\rho=0$.

{\bf iii}) Large-$\rho$ asymptotic form
\begin{eqnarray}
\eta (\rho,\phi) = \alpha\,\log\rho + o(1) \qquad \text{as}\quad
\rho \to\infty\,.
\end{eqnarray}

{\bf iv})  $\rho\to 0$ asymptotic form
\begin{eqnarray}\label{kakskasosi}
\eta(\rho,\phi) =
\begin{cases}
2l\,\log\rho  + O(1)
\qquad\qquad\quad \quad\quad \text{for} \quad |l|< {\textstyle\frac{1}{2}}\\
\pm\,\log\rho + O\big(\log(-\log\rho)\big)
\qquad\ \text{for} \quad l=\pm {\textstyle\frac{1}{2}}
\end{cases}
\ \ \qquad
\text{as}\quad
\rho\to 0 \ .
\end{eqnarray}

Unless specified otherwise, we will describe  the cone\ \eqref{cone} by the chart,
\bea\label{skklslk}
{\cal M}_z^{(0)}\ :\ \ \ \ \ \ \ 
\ -{\textstyle {\pi\over 2\alpha}}\leq \phi\leq {\textstyle {\pi\over 2\alpha}}\,, \qquad
0 \leq  \rho < \infty
\eea
with the rays $\phi=- {\pi\over 2\alpha}$ and $\phi= {\pi\over 2\alpha}$ identified.
We assume that solution satisfying
conditions {\bf i})-{\bf iv})  is unique, and hence respects all symmetries of
Eq.\eqref{shgz}. In particular,
\bea\label{saasksuq}
\eta(\rho,\,\phi)=\eta(\rho,-\phi)\ .
\eea

Starting from the asymptotic form \eqref{kakskasosi} one can develop
$z,\zb \to 0$ expansion of the form
\bea\label{zexp}
\eta& =&
l\ \log (z\zb )+\eta_0+\sum_{k=1}^\infty \gamma_k \ \big(\, z^{2\alpha k}+{\bar z}^{2\alpha k}\,\big)\\
&-&
{ s^{4\alpha}\, \re^{-2\eta_0}\over (1-2l)^2}\ \ (z \zb )^{1-2 l}+
{\re^{2\eta_0} \over (1+2l)^2}\ \ ( z{\bar z})^{1+2 l}+\ldots\ ,\nonumber
\eea
where $\eta_0$ and $\gamma_k$ are integration constants. It is easy to see
that the coefficients in all omitted terms in this expansion
are uniquely determined once these integration constants are given.
On the other hand, $\eta_0$ and $\gamma_k$ are not new parameters of the solution;
they have to be determined from consistency of this expansion with
the remaining conditions {\bf i})-{\bf iii}). We will give explicit form of the
constant $\eta_0$ in Section\,\ref{Sect2} below  (see Eqs.\eqref{hddagad},\,\eqref{kjsaaslkskl}).

The expansion \eqref{zexp} remains valid if we regard $z$ and $\zb$
as independent complex variables. For our analysis, the most
important message from \eqref{zexp} is that in the ``light-cone''
limit $\zb\to 0$ (with fixed $z$)
\bea\label{gammadef}
\eta \to l\,\log( z\zb)+\eta_0 + \gamma(z)\,,
\eea
where $\gamma(z)=\sum_{k=1}^\infty\gamma_k\,z^{2\alpha k}$ decays as
\bea\label{iskkslaal}
\gamma(z) \sim z^{2\alpha}
\eea
at small $z$.

Although the solution $\eta(\rho,\phi)$ is a single-valued functions on the cone
\eqref{cone}, the connection \eqref{ssoiso} is not. Instead, the linear problem
\eqref{lp} is invariant with respect to the operation
\begin{eqnarray}\label{kksksa}
{\hat\Omega}\,: \qquad \phi \to \phi + {\textstyle \frac{\pi}{\alpha}}\,, \quad \theta \to
\theta - {\textstyle \frac{\ri\pi}{\alpha}}\,,
\end{eqnarray}
involving the shift of the spectral parameter $\theta$. Another easily established
symmetry of this linear problem involves the operation
\begin{eqnarray}\label{gasfar}
{\hat\Pi}\,: \qquad \theta \to \theta - \ri\pi\,,
\end{eqnarray}
which transforms the connection \eqref{ssoiso} as
\begin{eqnarray}\label{sksksakls}
{\hat \Pi}[\, {\boldsymbol D}\,] = \sigma^3\, {\boldsymbol D}\,\sigma^3\,, \qquad
{\hat \Pi}[\, {\bar {\boldsymbol D}}\,] = \sigma^3\,{\bar {\boldsymbol D}}\,\sigma^3\,.
\end{eqnarray}
Motivated by  these mutually commuting  symmetries, we define two solutions
${\bf\Psi}_{\pm} = {\bf\Psi}_{\pm}(\rho,\phi\,|\,\theta)$ of the linear problem
\eqref{lp} uniquely specified by their asymptotic behavior
\bea\label{sasaa}
{\boldsymbol \Psi}_+\to {1\over \sqrt{\cos(\pi l)} }\
\begin{pmatrix}
0\\
\re^{( \ri \phi+\theta) l}
\end{pmatrix}\ ,\ \ \ \  \
{\boldsymbol \Psi}_-\to {1\over \sqrt{\cos(\pi l)} }\ \begin{pmatrix}
\re^{-(\theta+\ri\phi) l}\\
0
\end{pmatrix} \ \ \ \ \ {\rm as}\ \ \ \ \rho\to 0\, .
\eea
In writing \eqref{sasaa} we have assumed that $|l|<\frac{1}{2}$. In the analysis below we usually adopt
this limitation, and treat the case $|l|=\frac{1}{2}$ by continuity. Using Eqs.\eqref{kksksa}-\eqref{sasaa},
and the fact that at real $\theta$
\bea\label{asksksa}
{\bar {\boldsymbol D}}(\theta)=\sigma^1\, {\boldsymbol D}^*(-\theta)\, \sigma^1\ ,
\eea
where the star denotes complex 
conjugation and $\sigma^1=\sigma^++\sigma^-$, it is straightforward to establish
the following properties of these solutions:

\begin{itemize}
\item ${\boldsymbol \Psi}_\pm(\rho,\phi\,|\,\theta)$ are entire functions of
$\theta$ for arbitrary real $\phi$ and  $\rho>0$.
\item ${\hat\Omega}$-invariance:
\bea\label{ahaytatissu}
{\boldsymbol \Psi}_\pm \big(\rho,\phi+{\textstyle{\pi\over
2\alpha}}\,|\,\theta-{\textstyle{\ri\pi\over 2\alpha}}\big)= \
{\boldsymbol \Psi}_\pm(\rho ,\phi-{\textstyle{\pi\over 2\alpha}}
\, |\,\theta+{\textstyle{\ri\pi\over 2\alpha}})\ .
\eea
\item ${\hat\Pi}$-transformation
\bea\label{sssalksa} {\boldsymbol \Psi}_+ (\rho,\phi\, |\, \theta\pm
\ri\pi)&=&-\re^{\pm\ri\pi l}\   \sigma^3\
{\boldsymbol \Psi}_+(\rho,\phi\, |\,\theta)\,
,\\
{\boldsymbol \Psi}_- (\rho,\phi\, |\, \theta\pm \ri\pi)&=&\ \ \re^{\mp\ri\pi l}\ \sigma^3\
{\boldsymbol \Psi}_-(\rho,\phi\, |\, \theta)\ . \nonumber
\eea
\item Normalization condition
\bea\label{slslsa}
\det({\boldsymbol \Psi}_+,\,{\boldsymbol \Psi}_-)=-{1 \over \cos(\pi l)}\ .
\eea

Here and below $({\boldsymbol \Psi}_+,\,{\boldsymbol \Psi}_-)$ stands for the
$2\times 2$ matrix with the columns ${\boldsymbol \Psi}_+$ and ${\boldsymbol \Psi}_-$.
\item For real $\theta$
\bea\label{lslsak}
{\boldsymbol \Psi}^*_\pm(\rho,\phi\, |\, \theta)=\sigma^1\,
{\boldsymbol \Psi}_\mp(\rho,\phi\, |\, -\theta)\ ,
\eea
and
\bea\label{kkslsssal}
{\boldsymbol \Psi}_\pm^*(\rho,\,0\,|\,\theta)={\boldsymbol \Psi}_\pm(\rho,\,0\,|\,\theta)\  .\eea

\end{itemize}

Note also that for $l\to\pm \frac{1}{2}$:
\bea\label{mahah} \lim_{l\to\pm
\frac{1}{2}}\,\big[\,  \cos(\pi l)\, ({\boldsymbol \Psi}_+- {\boldsymbol \Psi}_-)\, \big]=0\ .
\eea

The above solutions ${\boldsymbol \Psi}_\pm$ are specified by their $\rho\to 0$ behavior
\eqref{sasaa}. On the other hand, 
at large $\rho$ the WKB analysis applies. Assuming that $\theta$ is real,
it is straightforward
to show that while generic solution of \eqref{lp} grows exponentially at $\rho\to \infty$, there
is a solution  which decays in the
wedge
\bea\label{slkslksa}
-{\textstyle {\pi\over 2(\alpha+1)}}\leq
\phi\leq
{\textstyle {\pi\over 2(\alpha+1)}}\ .
\eea
We denote this decaying solution as ${\boldsymbol \Xi}= {\boldsymbol \Xi}(\rho,\phi\,|\,\theta)$. It is
uniquely specified by the asymptotic condition
\bea\label{kklssaklsa}
{\boldsymbol \Xi}(\rho,\phi\,|\,\theta)\to {\boldsymbol {\cal E}}_-(\rho,\phi\,|\,\theta)
\ \ \ \ \ \ {\rm as}\ \ \ \ \rho\to+\infty\ ,
\eea
where ${\boldsymbol {\cal E}}_-$ is the shorthand for the decaying exponential
\bea\label{qwaasz}
{\boldsymbol {\cal E}}_-=
\begin{pmatrix}
 \re^{-{\ri \alpha\phi\over 2}} \\
-\re^{{\ri \alpha\phi\over 2}}
\end{pmatrix}
\ \exp\bigg[\, -{2\rho^{\alpha+1}\over \alpha+1}\
\cosh(\theta+\ri(\alpha+1)\phi\big)\, \bigg]\ .
\eea

Since ${\boldsymbol \Psi}_\pm$ form a basis  in the space of solutions of
linear problem\ \eqref{lp}, we have linear relation
\bea\label{slslsla}
{\boldsymbol \Xi}=Q_-(\theta)\ {\boldsymbol \Psi}_++Q_+(\theta)\
{\boldsymbol \Psi}_-\ ,
\eea
where the coefficients $Q_\pm (\theta)$ (of course independent of the variables
$\rho,\phi$) are functions of the spectral parameter $\theta$ as well as the parameter
$l$ (the last argument is temporarily omitted in the above notations). These coefficients
are to be related to the $Q$-function of the quantum sine-Gordon model \eqref{sg}.

As is well known, the matrix linear
problem \eqref{lp} can be reduced to second order linear
differential equations. One can write general solution of
\eqref{lp} as
\bea\label{ilslsa}
{\boldsymbol \Psi}=
\begin{pmatrix}
\re^{{\theta\over 2}}\ \re^{\eta\over 2}\ \psi\\
\re^{-{\eta\over 2}}\ \re^{-{\theta\over
2}}\,(\partial_z+\partial_z\eta)\, \psi
\end{pmatrix}=
\begin{pmatrix}
\re^{-{\eta\over 2}}\  \re^{\theta\over 2}\, ( \partial_{\zb}+
\partial_{\zb} \eta)\,   {\bar \psi}\\
\re^{{\eta\over 2}}\ \re^{-{\theta\over 2}}\ {\bar\psi}
\end{pmatrix}\ ,
\eea
where $\psi$ and ${\bar \psi}$ solve the equations
\bea\label{sksaksa}
&&\big[\, \partial_{z}^2-u(z,{\bar z})-\re^{2\theta}\ \ p(z)\, \big]\ \psi=0\ ,\\
&&\label{sksaksabar}\big[\, \partial^2_{\zb}-{\bar u}(z,{\bar z})-\re^{-2\theta}\,
{ p}({\bar z})\, \big]\ {\bar \psi}=0\ ,
\eea
with
\bea\label{sksklsalsa} u(z,{\bar
z})=(\partial_z\eta)^2-\partial_z^2\eta\ ,\ \ \ \ {\bar
u}(z,{\bar z})=({ \partial}_{\zb}\eta)^2-{ \partial}^2_{\zb}\eta\
.
\eea

This form is convenient for making connection to the analysis in
Ref.\cite{Bazhanov:1998wj} which emerges at small $s$ and large $\theta$.
Concentrating attention on \eqref{sksaksa},
consider first the light cone limit $\zb\to 0$ in which $\eta$ assumes the form
\eqref{gammadef}. After that, the limit $z
\sim s \to 0\,,\ \theta \to +\infty$ can be taken,
with the combinations
\bea\label{Pssksa}
x=\re^{{\theta\over 1+\alpha}}\ z\ ,\ \ \ \ \ \qquad
E= s^{2\alpha}\ \re^{{2\theta \alpha \over 1+\alpha}}\ ,
\eea
kept finite. Then the term $\gamma(z)$ in \eqref{gammadef} can be
dropped, and \eqref{sksaksa} reduces to the Schr${\rm \ddot o}$dinger equation
\bea\label{skksa}\label{schroedinger}
\bigg[\, -\partial_x^2+{l(l+1)\over x^2}+x^{2\alpha}\,\bigg]\
\psi=E\ \psi\ .
\eea
In this double limit the solutions ${\bf \Psi}_\pm$ assume asymptotic
form
\bea\label{kslslsa}
{\boldsymbol \Psi}_+ &\to& {\re^{{\alpha\theta (2l+1)\over
2(\alpha+1)}}\over (2l+1)\sqrt{\cos(\pi l)}}\ \ \ \
\begin{pmatrix}
(z\zb)^{l\over 2}\ \  \re^{{\theta\alpha \over 2 (\alpha+1)}}\  \psi_+ (x) \nonumber \\
\\
(z\zb)^{-{l\over 2}}\,  \re^{-{\alpha\theta \over 2
(\alpha+1)}} \ (\partial_x+{l\over x})\, \psi_+ (x)
\end{pmatrix}\ ,\\
{\boldsymbol \Psi}_- &\to& {\re^{-{\alpha\theta (2l+1)\over 2(\alpha+1)}}\over
\sqrt{\cos(\pi l)}}\ \ \ \
\begin{pmatrix}
(z\zb)^{l\over 2}\ \ \re^{{\theta\alpha \over 2 (\alpha+1)}}\  \psi_- (x)\\
\\
(z\zb)^{-{l\over 2}}\,  \re^{-{\alpha\theta \over 2
(\alpha+1)}}\ (\partial_x+{l\over x})\, \psi_- (x)
\end{pmatrix}\ ,
\eea
where $\psi_\pm$ are
unique solutions of the above Schr${\rm \ddot o}$dinger equation \eqref{schroedinger}
defined for $|l|< {1\over 2}$ by their $x\to 0$ behavior
\bea\label{laaksi}
\psi_+\to   x^{l+1}\,, \qquad
\psi_-\to   x^{-l}\ .\nonumber
\eea
The equation \eqref{slslsla} then reduces to
\bea\label{chipsi}
\chi = Q_{-}^{(\text{cft})}(\theta-\theta_s)\ \psi_+ + Q_{+}^{(\text{cft})}(\theta-\theta_s)\ \psi_-\ ,
\eea
where $\chi$ is the decaying solution of \eqref{schroedinger}, defined by the asymptotic
condition
\bea
\chi \to x^{-\frac{\alpha}{2}}\ \exp\Big[-\frac{x^{\alpha+1}}{\alpha+1}+O\big(x^{1-\alpha}\big)\, \Big]\ \ \ \ \
{\rm as}\ \ x\to +\infty\ ,
\eea
and $\theta_s=-(\alpha+1)\ \log(s)$.
The coefficients $Q_{\pm}^{\rm(cft)}(\theta)$ coincide with the $Q$-functions (the vacuum eigenvalues of the
operators ${\bf Q}_{\pm}(\theta)$ defined in
\cite{Bazhanov:1996dr,Bazhanov:1998dq}) of the left-moving\footnote{Of course
the $Q$-functions of the right-moving sector show up in the opposite double limit $z\to 0$ and
$\zb\sim s \gg z,\
\theta\to -\infty$ in \eqref{sksaksabar}.}
chiral sector of the CFT emerging in the massless case $\mu=0$ in \eqref{sg}. The proof of this statement
is given in Ref.\cite{Bazhanov:1998wj}.
Eq.\eqref{slslsla} generalizes \eqref{chipsi} to the massive case $\mu \neq 0$.

\section{\label{Sect2} $Q$-function}

\subsection{\label{baseprop}Basic properties}

The following properties are established by arguments nearly identical to those
presented in Ref.\cite{Bazhanov:1998wj} (For completeness we sketch the derivations
in  Appendix A)

\begin{itemize}

\item
$Q_\pm(\theta)$ are entire,  quasiperiodic   functions of
$\theta$,
\bea\label{slslsaiaua} Q_\pm\big(\theta+{\textstyle
{\ri \pi\over 2}}+{\textstyle {\ri \pi\over 2\alpha}}\big)=\re^{\pm\ri\pi (l+{1\over 2})}\
\ Q_\pm\big(\theta-{\textstyle {\ri \pi\over
2}}-{\textstyle {\ri \pi\over 2\alpha}}\big)\  .
\eea

\item  At real $\theta$
\bea\label{sklsaksklsa}
Q^*_\pm (\theta)=Q_\pm (\theta)\ ,\ \ \ \ \
Q_\pm (\theta)=Q_\mp (-\theta)\ .
\eea

\item $Q_\pm (\theta)$ satisfy the Quantum Wronskian relation
\bea\label{oosaopsa} Q_+(\theta+ {\textstyle {\ri\pi\over
2\alpha}})Q_-(\theta- {\textstyle {\ri \pi\over 2\alpha}})-
Q_+(\theta- {\textstyle {\ri\pi\over 2\alpha}})Q_-(\theta+
{\textstyle {\ri \pi\over 2\alpha}})=-2\ri\ \cos(\pi l)\ .
\eea

\item
\bea\label{sklsklsalas}
\lim_{l\to\pm \frac{1}{2}}Q_{+}(\theta)=\lim_{l\to \pm \frac{1}{2}}Q_{-}(\theta)\ .
\eea

\end{itemize}

We now define the function $Q(\theta,k)$ of complex $\theta$ and real $k$ as follows.
For $k\in (- {1\over 2},\,  {1\over 2})\backslash 0$ we set
\bea\label{ksklsalksa}
Q(\theta, k):=
\begin{cases}
Q_+(\theta)\big|_{l=2 k-{1\over 2}}\ \ \ &{\rm for}\ \  0<k<{\textstyle {1\over 2}}\\
Q_-(\theta)\big|_{l=-2 k-{1\over 2}}\ \ \ &{\rm for}\ \
-{\textstyle {1\over 2}}<k<0
\end{cases}\ .
\eea
Due to the property \eqref{sklsklsalas} this definition extends to $k=0, \ k=\pm \frac{1}{2}$
by continuity, and then admits continuous\footnote{Our analysis in subsections\,\ref{Zeroes},\,\ref{DDV}
suggests
that $Q(\theta,k)$ is in fact analytic at all real $k$.} periodic extension to all real $k$ by
\bea\label{asossu}
Q(\theta, k)=Q(\theta, k+1)\ .
\eea
Thus defined, $Q(\theta, k)$ is entire function of
$\theta$, satisfying
\bea\label{slaua}
Q\big(\theta+{\textstyle
{\ri \pi(\alpha+1)\over \alpha}},\, k\big)= \re^{2\pi\ri k}\
Q\big(\theta,\, k)\ ,
\eea
\bea\label{qreality}
Q^*(\theta,k)=Q(\theta^*,k)\ ,\ \ \
Q(-\theta,k)=Q(\theta,-k)\ ,
\eea
and
\bea\label{oosaussyopsaisu}
Q(\theta+ {\textstyle
{\ri\pi\over 2\alpha}}, k)\, Q(\theta- {\textstyle {\ri \pi\over
2\alpha}}, -k)- Q(\theta- {\textstyle {\ri\pi\over 2\alpha}},k)\,
Q(\theta+ {\textstyle {\ri \pi\over 2\alpha}},-k)= -2\ri \,
\sin(2\pi k)\ .
\eea

To fix the function $Q(\theta,k)$ uniquely, we will need two additional
analytic properties. One concerns with the asymptotic behavior at
$|\Re e\,(\theta)|\to\infty$, another determines the general pattern
of zeros of this function in the complex $\theta$-plane.
Due to the quasi-periodicity
\eqref{slaua} one can concentrate attention on the strip
\bea\label{oskskskjsa}
H\ :\ \ \big|\Im m\, (\theta)\big|\leq
{\textstyle{\pi(\alpha+1)\over 2\alpha}}\ .
\eea Define also
\bea\label{amaayay}
H_+\ :\ \ \  0< \Im m (\theta) <
{\textstyle{\pi(1+\alpha)\over \alpha}}\ ;\ \ \ \ \ \ H_-\ :\ \ \
-{\textstyle{\pi(1+\alpha)\over \alpha}} < \Im m\,\theta<0\ .
\eea
Then

\begin{itemize}

\item For real $k$ and  $\alpha^{-1}\not =1,\,3,\,5\ldots$\ ,
\bea\label{jsajksajka}
Q&\to&\re^{ \pm \ri\pi  k}\,   {\mathfrak S}^{1\over 2}
\ \exp\bigg[\,{r \, \re^{\theta\mp {\ri\pi(1+\alpha)\over
2\alpha}}\over 4\cos({\pi\over 2\alpha})}\,\bigg]\ \ \ \ \ \
\Big(\, \theta\in H_\pm\, , \ \Re e\,(\theta)\to +\infty\,\Big)\\
Q&\to&\re^{ \pm \ri\pi k}\, { \mathfrak S}^{-{1\over 2}} \ \exp\bigg[\,{r \,
\re^{-\theta\pm {\ri\pi(1+\alpha)\over 2\alpha}}\over
4\cos({\pi\over 2\alpha})}\,\bigg]\ \ \Big(\, \theta\in H_\pm\, ,
\ \Re e\,(\theta)\to -\infty\,\Big)\ .\nonumber
\eea
Here
\bea\label{sksksa} r=B\ s^{1+\alpha}\
,\ \ \ \ \ \ \ \ \ \ B={2\sqrt{\pi}\Gamma(1+{1\over 2\alpha})\over
\Gamma({3\over 2}+{1\over 2\alpha})}\ , \eea
and
${\mathfrak S}$ is related to the constant $\eta_0$  in Eq.\,\eqref{zexp}
as follows:
 \bea\label{hddagad}
  {\mathfrak S}= {\Gamma(2k)\over \Gamma(1-2 k)}\ 2^{4k-1}\
 \re^{\eta_0}\ \ \ \ \ \ \ \ \ \  \ \ (0\leq k<{\textstyle\frac{1}{2}})\ .
 \eea
${\mathfrak S}={\mathfrak S}(k)$ is  a real  function of  real variable  $k$,
such that \bea\label{alsksjasuu}
{\mathfrak S}(k)\, {\mathfrak  S}(-k)=1\ , \ \ \ \ \ {\mathfrak S} (k+1)={\mathfrak S} (k)\ .
\eea

\item For any real $k$,  all  the  zeros of $Q(\theta,k)$ in the strip
$H$ are real, simple, and accumulate towards $\theta\to\pm\infty$. Let
\bea\label{sksksak}
\epsilon(\theta) =\ri\  \log\left[\,\frac{Q\left(\theta+\frac{\ri\pi}{\alpha},\,k\right)}
{Q\left(\theta-\frac{\ri\pi}{\alpha},\,k\right)}\,\right]\ ,
\eea
where the branch of the log is fixed by the condition
\bea
\epsilon(\theta)-\frac{r\,\re^\theta}{2\cos(\frac{\pi}{2\alpha})}\ \to\ -2\pi \,k \qquad
\text{for} \quad \Re e (\theta) \to +\infty\quad \text{and}\quad
|\Im m (\theta)|<{\textstyle \frac{\pi}{2}}\ .
\eea
Then the zeros $\theta_n$ can be labeled by consecutive integers $n=0,\, \pm 1,\, \pm 2,\, \ldots\,$, so
that $\theta_n < \theta_{n+1}$, and
\bea\label{kksklsa}
\epsilon(\theta_n) = \pi\  (2n+1)\ .
\eea

\end{itemize}

\noindent
The asymptotics \eqref{jsajksajka} can be established
through straightforward WKB analyses  of the linear differential
equations\ \eqref{lp}. The second property is derived in the next subsection.

\bigskip

\subsection{\label{Zeroes}Zeroes}

Consider first the limit of small $s$.  Due to the
relations\ \eqref{asossu} and\ \eqref{qreality} we
can assume, without loose of generality, that
$0\leq  k \leq \frac{1}{2}$.
In this case the problem reduces to
Schr${\rm \ddot o}$dinger  equation \eqref{schroedinger}, and the limiting pattern of zeros
of $Q(\theta,k)$ can be read out from the pattern of zeros of $Q_{\pm}^{\rm (cft)}(\theta)$,
which is relatively well understood \cite{Bazhanov:1996dr,Bazhanov:1998wj}.
Simple analysis shows
that  in this case the zeros $\{\theta_n\}$ split into two widely separated groups,
the ``positive'' zeros $\{\theta_n\}|_{n=0}^\infty$, and the ``negative'' zeros
$\{\theta_{-n-1}\}|_{n=0}^\infty$.
The limiting behavior of the positive zeros follows from
Eqs.\eqref{Pssksa}-\eqref{laaksi},
\bea\label{iujskajs}
s^{2\alpha}\ \re^{2\alpha\theta_n\over\alpha+1}\ \to \ {\cal E}_n(l)\ ,
\eea
where $\big\{ {\cal E}_n(l)\big\}\big|_{n=0}^\infty$
are eigenvalues of   Schr${\rm \ddot o}$dinger  operator \eqref{schroedinger}
associated with the spectral problem
\bea\label{jjsjsajsa}
\psi&\to& x^{l+1}\ \ \ {\rm as}\ \ \ x\to 0\\
\psi&\to& 0\     \ \  \ \ \ \  {\rm as}\ \  \
x\to+\infty\ , \nonumber
\eea
with $l=2k-\frac{1}{2}$.
If  $l>-\frac{1}{2}$
the  spectral problem \eqref{schroedinger}, \eqref{jjsjsajsa}
corresponds to  self-conjugated Schr${\rm \ddot o}$dinger  operator,
and the associated
eigenvalues are non-degenerate and positive.
In fact, it is possible to show that
$\big\{{\cal E}_n(l)\big\}_{n=0}^\infty$
is a family of meromorphic functions
of  complex  $l$. Furthermore,
they  are  analytic in the strip $\Re e\,(l)\geq -\frac{3}{2}$ and
 ${\cal E}_n(l)>0$
for real  $l>-\frac{3}{2}$.
At  $l=-\frac{3}{2}$ we have ${\cal E}_0(-\frac{3}{2})=0$, while
all higher ${\cal E}_n(l)$ $(n=1,\,2\ldots)$ remain positive.
Therefore at $s\to 0$ all the positive zeros are real,  simple
and have the limiting behavior \eqref{jjsjsajsa}
for any
$ k\in \big[ 0,\, \frac{1}{2}\big]$, including  the ends of  the  segment.

The limiting behavior of the negative zeros
 is  slightly  more  delicate, but still
can be described by  equation   similar to  \eqref{iujskajs},
\bea\label{skajs}
s^{2\alpha}\ \re^{-{2\alpha\theta_{-n-1}\over\alpha+1}}\ \to {\cal E}_n(-1-l)\ .
\eea
Here the argument $-1-l$ belongs to the segment $\big[-\frac{3}{2},-\frac{1}{2}\big]$, where
${\cal E}_n(-1-l)$ are understood as the  analytic
continuations from the domain $\Re e(l)>\frac{1}{2}$.
It is important that all ${\cal E}_n(-1-l)$ remain strictly  positive, except
for ${\cal E}_0(-1-l)$, which turns to zero at $l=\frac{1}{2}$.   Thus
the limiting behavior of the negative zeros
is similar to the positive ones, i.e.
they  are also  real and simple.
If $s$ is small but finite, at  $k\to \frac{1}{2}$ the single zero  $\theta_{-1}$
leaves the group of  negative zeroes and  travels  toward the positive ones.
In the process, $\theta_{-1}$ remains real for  any
$0\leq  k< \frac{1}{2}$, since $Q(\theta,\, k)$ is a real analytic function.
For $k=\frac{1}{2}$ and arbitrary  $s>0$ we have $\theta_{-1}\sim 1$, and hence it
still formally  obeys Eq.\eqref{skajs}.

By continuity in $s$ (which we assume) this pattern of well separated
groups of positive and negative zeros persists for sufficiently small
values of this parameter. While exact locations of the zeros change with
$s$, no additional zeros can be generated, since this would violate the
asymptotics form \eqref{jsajksajka}, which is valid for any $s$. One can
define two positive and non-degenerate
sets of numbers $\big\{E_n(\pm k)\big\}|_{n=0}^\infty$ by
\bea\label{slsslsa}
\re^{2\alpha\theta_n\over \alpha+1}=
\begin{cases}
\ s^{-2\alpha}\ E_n(k)\ ,\ \ \ \ \ \ \ \ \ \ &n\geq 0\\
\ s^{2\alpha}\ E^{-1}_{-n-1}(-k)\ ,\ \ \ &n<0
\end{cases}\ .
\eea
In the domain $0\leq k\leq   \frac{1}{2}$, at $s\to 0$
the sets  $\big\{E_n(+ k)\big\}|_{n=0}^\infty$ and $\big\{E_n(- k)\big\}|_{n=0}^\infty$
converge to $\big\{{\cal E}_n\big(2k-\frac{1}{2}\big)\big\}|_{n=0}^\infty$ and
$\big\{{\cal E}_n(-2k-\frac{1}{2}\big)\big\}|_{n=0}^\infty$,
respectively. But at any $s$ we have
\bea\label{slsls} E_n(\pm k)\to
\big(\,{\textstyle{2\pi\over B}}\,  (2 n\pm 2k+1)\,\big)^{2\alpha\over \alpha+1}\ \ \big(\,
1+o(1)\, \big)\ \ \    \ {\rm as}\ \  \ n\to+\infty\, ,
\eea
where the constant $B$  is given by\ \eqref{sksksa}. This follows from the asymptotic
\eqref{jsajksajka}.
The entire function $Q(\theta,k)$  can be represented as
the product over its zeros. Thus, for $\alpha>1$ one can simply write
\bea\label{isuslssls}
Q(\theta,k)={\mathfrak C}(k)\ \re^{2k \alpha \theta\over \alpha+1}\,
\prod_{n=0}^{\infty} \bigg(\,1-{ s^{2\alpha}\, \re^{{2\alpha \theta \over
\alpha+1}}\over E_n(k)}\,\bigg)\,
\bigg(\,1-{s^{2\alpha}\, \re^{-{2\alpha\theta\over \alpha+1}}\over
E_n(-k)}\,\bigg)\ ,
\eea
where ${\mathfrak C} (k)$ is some $\theta$-independent factor. For
$0<\alpha\leq 1$
one has to include Weierstrass prime multiplier to make the product convergent.

Eqs.\eqref{asossu}-\eqref{qreality} imply the relations
\bea\label{sksksklsa}
E_n(k+1)=E_{n+1}(k)\ ,\ \ \ \
E_0(-k-1)\,E_0(k)=s^{4\alpha}
\eea
and
\bea\label{hytr}
{\mathfrak C}(k)=- s^{-2\alpha}\, E_0(k)\ 
{\mathfrak C} (k+1)\, ,\ \ \ \ \ \ \ \ \ \ {\mathfrak C} (k)={\mathfrak C}(-k)\ .
\eea
The overall normalization
factor ${\mathfrak C} $ in \eqref{isuslssls} and the ``reflection $S$-matrix''
\eqref{hddagad},\,\eqref{alsksjasuu}  also can be represented
as the convergent products
\bea\label{klsaksklsa}
&&
{\mathfrak C}(k)=(-1)^{ [k]}\
2^{\alpha\over\alpha+1}\ \prod_{n=0}^{\infty}\sqrt{ E_n(k)\,
E_n(-k)}\  \  \big({\textstyle{2\pi\over B}}\, (2n+1)\,\big)^{-{2\alpha\over \alpha+1}} \
\eea
and
\bea\label{sssaisai} 
&& {\mathfrak S}  (k)=\bigg({r\re^{\gamma_E}\over 4\pi}\bigg)^{-{4k\alpha\over\alpha+1}}\  \
 \prod_{n=0}^{\infty}
{E_n(-k)\over E_n(k)}\ \ \re^{4\, k\, \alpha\over (\alpha+1)(n+1)}\ ,
\eea
where $[k]$ is an integer part of real number $k$,
$\gamma_E$ is the Euler constant, and we use $r$ from \eqref{sksksa}.

The  Quantum Wronskian relation \ \eqref{oosaussyopsaisu} shows that at any real zero $\theta_n$
\bea\label{kjsksksa}
\re^{-\ri \epsilon (\theta_n)}=-1\ ,
\eea
where $\epsilon(\theta)$ is defined in \eqref{sksksak}. Hence,
${\epsilon(\theta_n)\over\ri\pi}$ is   an odd integer number which, by continuity, cannot change
when one changes $s$.  Therefore these numbers can be extracted
from the limit $s\to 0$ (see e.g. Appendix A in Ref.\cite{Bazhanov:2003ni}). 
By this argument, Eq.\eqref{kksklsa} is valid for any finite $s$.

Since $\tau(\theta)={\epsilon(\theta)\over 2\pi}+ k$ is real valued, monotonically  increasing
function of the real variable $\theta$, one  can introduce the inverse function
$\theta(\tau)$,
which has the properties
\bea\label{slksklsak}
{\theta}(\tau)&=&-{\theta}(-\tau)\  \ \ \ \ \ \ \ \ \ \ \ \   \big(\,\Im  m(\tau)=0\,\big) \\
{\theta}(\tau)&\to&\log\Big( { 2\pi\tau\over r}\Big)\ \ \ \ {\rm as}\ \ \ \
\tau\to+\infty\ .\nonumber
\eea
Then,
\bea\label{kss}
E_0(k)=s^{2\alpha}\ \exp\Big[\, {\textstyle{2\alpha\over \alpha+1}}\
\theta(2 k+1)\,\Big]\ ,
\eea
and
\bea\label{aksskaj}
E_n(k)=E_0(k+n)\ ,\ \ \ \ \ \ E_n(-k)={s^{4\alpha}\over E_0(k-n-1)}\ \ \ \ 
\qquad (n=0,\,1\ldots)\ .\nonumber
\eea

\subsection{\label{DDV}Destri-De Vega equation}

The quasiperiodic entire function $Q(\theta,k)$ is completely
determined by its zeros $\theta_n$ in the strip \eqref{oskskskjsa},
and the asymptotic condition \eqref{jsajksajka}. On the other hand,
the positions of the zeros $\theta_n$ are restricted by the equation
\eqref{kksklsa}. Mathematically, the problem of reconstructing the
function $Q(\theta,k)$ from this data has emerged long ago in the context
of the analytic Bethe ansatz \cite{Baxter:1971, Baxter:1982, Sklyanin:1978, Reshetikhin:83}.
For the sine-Gordon model \eqref{sg},
the problem was solved by Destri and
De Vega \cite{Destri:1992qk}, who have reduced it to a single
complex integral equation, the  DDV equation.

Starting with the equations
 \eqref{isuslssls}, \eqref{kksklsa}, \eqref{jsajksajka}, and
following the steps
outlined in \cite{Destri:1992qk,Destri:1997yz,Klumper:1991}, one can derive the equation\footnote{To be
precise, the derivation in \cite{Destri:1992qk, Destri:1997yz} goes through the analysis
of discretized version of the sine-Gordon system, where the analytic properties
of the counting function are somewhat different, with the continuous limit
taken at the end. Only minor adjustment of  the DDV arguments are needed in the
present context, see \cite{Bazhanov:1996dr}.}
\bea\label{ksalaks}
 2\int_{-\infty}^{\infty} \rd \theta'\,
\Im m\, \Big[\, \log\Big(\, 1+\re^{-\ri
\epsilon(\theta'-\ri 0)}\,\Big)\, \Big]\, G(\theta-\theta')= r\,
\sinh(\theta)- 2\pi  k-\epsilon(\theta)\ ,
\eea
where
\bea\label{ajkaskas}
G(\theta)=\int_{-\infty}^{\infty}{\rd\nu\over 2 \pi} \,
{\re^{\ri\theta\nu}\,
\sinh\big({\pi\nu(1-\alpha)\over2\alpha}\big)\over
2\cosh\big({\pi\nu\over2}\big)\,
\sinh\big({\pi\nu\over2\alpha}\big)}\ ,
\eea
and $r$ is given by\ \eqref{sksksa}.

Equation \eqref{ksalaks} coincides exactly with the DDV equation for the sine-Gordon
model \eqref{sg} if one sets
\bea
\alpha = \beta^{-2}-1\,, \qquad r=MR\,,
\eea
where \cite{Zamolodchikov:1995xk}
\bea
M =\frac{2\, \Gamma\big(\frac{\beta^2}{2-2\beta^2}\big)}{\sqrt{\pi} \Gamma\big(\frac{1}{2-2\beta^2}\big)}\
\bigg[\frac{\pi\mu \Gamma(1-\beta^2)}{\Gamma(\beta^2)}\bigg]^{\frac{1}{2-2\beta^2}}
\eea
is the sine-Gordon soliton mass, and identifies $k$ with the sine-Gordon quasi-momentum.
We conclude that the function $Q(\theta,k)$, defined in terms of the coefficients in \eqref{slslsla} as in
 Eq.\eqref{ksklsalksa}, coincides with the $Q$-function \cite{Baxter:1971,Baxter:1982,
Sklyanin:1985, Sklyanin:1991ss,Smirnov:1998kv}
of the quantum sine-Gordon model.

Few remarks about the equation \eqref{ksalaks} are worth making.
Note that at $\alpha=1$ the kernel $G(\theta)$ vanishes at all
$\theta$. In this case the solution of the DDV equation takes
simple form
\bea\label{trsksksa}
\epsilon(\theta)|_{\alpha=1}=\pi
s^2\, \sinh( \theta)-2\pi k\ .
\eea
One can use  \eqref{trsksksa}
to obtain   explicit form for $E_0(k)$\ \eqref{kss} in this
case,
\bea\label{uayatat}
E_0(k)|_{\alpha=1}=    2k+1+ \sqrt{
(2k+1)^2+ s^4}\ .
\eea
For general $\alpha>1$,  Eqs.\eqref{sksksak} and\
\eqref{jsajksajka} allow one to reconstruct the $Q$-function.
In particular for $\Im  m (\theta)={\pi (\alpha+1)\over 2\alpha}$
(i.e.  at the middle of the strip $H_+$\, \eqref{amaayay})
$Q(\theta,k)$ is  expressed in  terms of the solution of  DDV
equation as follows:
\bea\label{sksklsskla}
&&\log\,
Q\big(\theta+{\textstyle{\ri\pi (\alpha+1)\over
2\alpha}},\,k\big)= {r\, \cosh(\theta)\over 2\cos({\pi\over
2\alpha})}+\ri\pi k+
{\textstyle{1\over 2}}\ \log ( {\mathfrak S} )\\
&&\ \ \ \ +2\ri\, \int_{-\infty}^{\infty}  \rd\theta'\ \Im
m\bigg[\, \log\Big(1+\re^{-\ri \epsilon(\theta'-\ri 0)}\, \Big)
\, F(\theta'-\theta-\ri 0)\,\bigg]\ \ \ \ \ \ \ \ \big(\, \Im
m\,(\theta)=0\,\big)\ .\nonumber
\eea
Here,
\bea\label{sossao}
F(\theta)=\int_{-\infty}^{\infty} {\rd\nu\over 2\pi}\
{\re^{\ri\nu\theta}\over 4\cosh\big({\pi\nu\over 2}\big)\,
\sinh\big({\pi(\nu-\ri 0)\over 2\alpha}\big)}\ , \eea and
\bea\label{kjsaaslkskl} \log ( {\mathfrak S} )= \alpha\,
\int_{-\infty}^{\infty} {\rd\theta\over \pi}\, \Im m \Big[\,
\log\Big(1+\re^{-\ri \epsilon(\theta-\ri 0)}\Big)\, \Big]\ .
\eea
Once $\epsilon(\theta)$ is determined at real values of $\theta$, Eq.\eqref{sksklsskla}
can be modified to yield the
$Q$-function in the whole strip  $H_+$, and then by
\eqref{slslsaiaua} in the whole complex plane. For $0<\alpha\leq 1$ Eq.\eqref{sksklsskla} requires
minor modification.

\subsection{Large-$\theta$ asymptotic expansion and local IM }

Using \eqref{sksklsskla} it is straightforward to obtain
the large-$\theta$  asymptotic expansion of
the $Q$-function.
For $\big|\Im m\,(\theta)\big|<{\pi (\alpha+1)\over 2\alpha}$ we have
\bea\label{ajahya}
&&\log\, Q\big(\theta+{\textstyle {\ri\pi (\alpha+1)\over 2\alpha}}\, \big|\, k\big)\ \sim\
{r\,\re^\theta\over 4 \cos({\pi\over 2\alpha}) }+
\ri \pi k+{\textstyle {1\over 2}}\  \log( {\mathfrak S}) -\\
&&\ \ \ \sum_{n=1}^{\infty}\   {\mathfrak I}_{2n-1}
\ \ \re^{-(2n-1)\theta }+
\sum_{n=1}^{\infty}\
 {\mathfrak G}_n
 \ \  \re^{-2\alpha n\theta}\nonumber
\eea
as $\Re e\, (\theta)\to +\infty$, and
\bea\label{jsaskajsksa}
&&\log\, Q\big(\theta+{\textstyle {\ri\pi (\alpha+1)\over 2\alpha}}\, \big|\, k\big)\ \sim\
{r\, \re^\theta\over 4 \cos({\pi\over 2\alpha}) }+
\ri \pi k-{\textstyle {1\over 2}}\  \log({\mathfrak S})- \\
&&\ \ \ \sum_{n=1}^{\infty}\    {\bar {\mathfrak I}}_{2n-1}\ \ \re^{(2n-1)\theta } +
\sum_{n=1}^{\infty}
 {\bar {\mathfrak G}}_n\ \   \re^{2\alpha n \theta }\ \nonumber
\eea
as $\Re e\, (\theta)\to -\infty$.
Here we use the notations,
\bea\label{ssklslkas}
{\mathfrak I}_{2n-1}&=&-{r\, \delta_{n,1}\over 4\cos({\pi\over 2\alpha})}+
{ (-1)^{n+1}\over\sin({\pi (2n-1)\over\alpha})}\
\int_{-\infty}^{\infty}
{\rd\theta\over \pi}\,  \Im m
\Big[\, \re^{ (2 n-1)(\theta-\ri 0) }\,
\log\big(1+\re^{-\ri \epsilon(\theta-\ri 0) }\big)\, \Big]\nonumber\\
{\mathfrak G}_{n}&=& {\alpha\ (-1)^{n}\over\cos(\pi\alpha n)}\ \int_{-\infty}^{\infty}
{\rd\theta\over \pi}\ \Im m
\Big[\, \re^{2\alpha n (\theta-\ri 0) }\,
\log\big(1+\re^{-\epsilon(\theta-\ri 0)}\big)\, \Big]\ ,
\eea
and
\bea\label{usyssklslkas}
{\bar {\mathfrak I}}_{2n-1}&=&
-{r\, \delta_{n,1}\over 4\cos({\pi\over 2\alpha})}-
{ (-1)^{n+1}\over\sin({\pi (2n-1)\over\alpha})}\
\int_{-\infty}^{\infty}
{\rd\theta\over \pi}\,  \Im m
\Big[\, \re^{-(2 n-1) (\theta-\ri 0) }\,
\log\big(1+\re^{-\ri \epsilon(\theta-\ri 0) }\big)\, \Big]\nonumber\\
{\bar {\mathfrak G}}_{n}&=&- {\alpha \ (-1)^{n} \over\cos(\pi\alpha n)}\ \int_{-\infty}^{\infty}
{\rd\theta\over \pi}\ \Im m
\Big[\, \re^{-2\alpha n (\theta-\ri 0) }\,
\log\big(1+\re^{-\epsilon(\theta-\ri 0)}\big)\, \Big]\ .
\eea

\begin{figure}[!ht]
\centering
\includegraphics[width=6  cm]{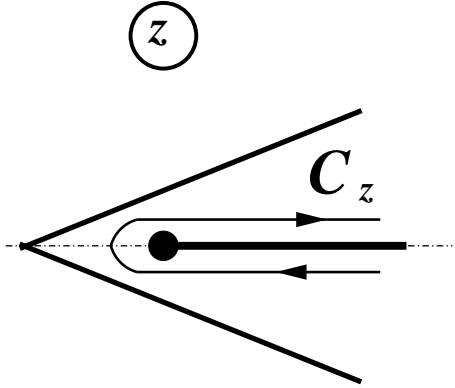}
\caption{The chart $\mathcal{M}_{z}^{(0)}$, and the integration contour $C_z$ in
Eq.\eqref{ksiauyalsklsa}.}
\label{fig1}
\end{figure}

On the other hand, the asymptotic expansions
\eqref{ajahya} and  \eqref{jsaskajsksa}
can be obtained directly from the WKB expansion
for  the linear  differential equations\ \eqref{lp}. To simplify calculations, it is convenient
to trade the variable $z$ to
\bea\label{uyslaskasa}
 w=\int \rd z\  \sqrt{p(z)}\ ,
\eea
and similarly for $\wb$. As is well known, this transformation brings MShG equation
\ \eqref{shgz} to the conventional Sinh-Gordon (ShG) equation
\bea\label{luuausay}
\partial_w{ \partial}_{\bar w}{\hat\eta}-\re^{2{\hat\eta}}+\re^{-2{\hat\eta}}=0
\eea
for
\bea\label{sksasa}
{\hat\eta}= \eta-{\textstyle \frac{1}{4}}\ \log \left(p  {\bar p}\right)\ ,
\eea
with
\bea\label{ajksasks}
p=p(z)\,,\ \ \ \  \ \ \ \ \ \ \ \ \ {\bar p}=p({\bar z})\ .
\eea
In what follows, we will assume that $\wb=w^*$, and choose the branch of $\sqrt{p(z)}$ in
\eqref{uyslaskasa} which is positive at the upper edge of the branch cut in Fig.\ref{fig1}.

Using the standard  technique\ \cite{Faddeev:1987ph} for the large-$\theta$ expansion, one can find
explicit expressions for the coefficients ${\mathfrak I}_{2n+1}$, 
 ${\mathfrak  G}_{n}$ and ${\bar {\mathfrak I}}_{2n+1},\
{\bar {\mathfrak  G}}_{n}$ as the functionals of ${\hat \eta}$. The coefficients ${\mathfrak I}_{2n+1}$
and  ${\bar {\mathfrak I}}_{2n+1}$ appear as local functionals, i.e. the integrals of local densities,
\bea\label{uissksksks}
{\mathfrak I}_{2n-1}={1
\over 2 (2n-1) \sin({\pi (2n-1)\over\alpha}) }\
\int_{C_w} \Big[\,\rd w\ {\hat P}_{2n}+
\rd {\bar w}\ {\hat  R}_{2n-2}\,\Big]
\eea
and
\bea\label{uisksks}
{\bar {\mathfrak I}}_{2n-1}={1
\over 2 (2n-1) \sin({\pi (2n-1)\over\alpha})}\  
 \int_{{\bar  C}_w}\Big[\, \rd {\bar w}\ {\hat {\bar P}}_{2n}+
\rd { w}\ {\hat {\bar  R}}_{2n-2}\, \Big]\ .
\eea
The integration contour $C_w$ here
is a  $w$-image of   the
contour $C_z$ in the $z$-plane shown on Fig.\ref{fig1}, while  ${\bar C}_w=C^*_w$.
The functions $({\hat  P}_{2n},\,{\hat R} _{2n-2})$ and
$( {\hat {\bar P}}_{2n}, {\hat {\bar R}}_{2n-2})=
({\hat P}^*_{2n},\,{\hat { R}}^*_{2n-2})$
are  conventional  tensor  densities of the local IM
for the ShG equation\ \eqref{luuausay}, satisfying the continuity equations
\bea\label{sksaklsalas}
\partial_{\bar w}{\hat P}_{2n}=\partial_w {\hat  R}_{2n-2}\ .
\eea
They can be obtained in  explicit form as follows. Let
\bea\label{sshjaqreew}
{\hat u}&=&(\partial_w{\hat \eta})^{2}-\partial^2_w{\hat \eta}\, ,\ \ \ \ \ \ \ \ \
{\hat v}=(\partial_w{\hat \eta})^{2}+\partial^2_w{\hat \eta}\ .
\eea
Then
\bea\label{ystksksksa}
{\hat P}_{2n} &=&U_{n}[\, {\hat u}\, ]\ ,\ \ \ \ \ \ \ \ {\hat R }_{2n-2}=
\re^{-2{\hat\eta}}\  U_{n-1}[\, {\hat v}\,]-
 \delta_{n,1}
\ ,\nonumber
\eea
where $U_{n}[\,  {\hat u}\, ]$ are  
homogeneous $({\rm grade}({\hat u})=2, \ {\rm grade}(\partial )=1,\
{\rm grade}(U_n)=2n )$
differential polynomials  in ${\hat u}$ of the degree 
$n$  (known as the Gel'fand-Dikii polynomials \cite{Gelfand}),
\bea\label{iusksksak}
U_{n}[\,{\hat u}\,]= {\hat \Lambda}^n\cdot 1\ .
\eea
Here
\bea\label{reqssksa}
{\hat \Lambda}=-{\textstyle {1\over 4}}\ 
\partial^2+ {\hat u}-{\textstyle {1\over 2}}\  {\partial}^{-1}\  {\hat u}'  \, ,
\eea
and prime stands for the derivative. Thus,
\bea\label{rwesksksa}
U_0[\, {\hat u}\,]&=& 1\ ,\nonumber\\
U_1[\, {\hat u} \,]&=& {\textstyle {1\over 2}} {\hat u}\ ,\\
U_{2}[\,{\hat u} \,]&=& {\textstyle {3\over 8}}\, 
{\hat u}^2-{\textstyle {1\over 8}}\, {\hat u}''\ , \nonumber\\
U_{3}[\, {\hat u}\,]&=& {\textstyle {5\over 16}}\, {\hat u}^3-{\textstyle {5\over 32}}\, ( {\hat u} ')^2-
{\textstyle {5\over 16}}\,{\hat u}\, ''+
{\textstyle {1\over 32}}\ {\hat u}''''\ , \nonumber\\
U_{n}[\,{\hat u}\,]&=&{\textstyle {
\Gamma(n+{1\over 2})\over n!\sqrt{\pi}}}\ \ {\hat u}^n+\ldots\ ,\nonumber
\eea
where the last line shows overall normalization of the polynomials.

Of course, it is straightforward to rewrite the integrals \eqref{uissksksks}, \eqref{uisksks} in terms of
the original variables $z,\zb$, in which the integration contour is just $C_z$ in Fig.\ref{fig1}. We have
\bea\label{usysskskls}
&&\rd w=\sqrt{p(z)}\ \rd z\ ,\nonumber\\
&&{\hat u}= p^{-1}\ \bigg[\,  u+{4 p p'' -5 p'^2\over 16 p^2}\, \bigg]\ ,
\eea
where $p=p(z)$, $u=(\partial_z\eta)^2-\partial^2_z\eta$, so that
\bea
&&{\hat P}_{2n}[\,{\hat u}\,]= p^{-n}\ P_{2n}[\,u\,]+\ldots\  .
\eea
For example
\bea\label{ksiauyalsklsa}
{\mathfrak I}_{1}\ \sin\big({\textstyle{\pi\over 2\alpha}}\big)=-{\pi (2\alpha-1)\over 12 r (\alpha+1)}+{1
\over 2 }\
\int_{C_z} \Big[\,\rd z\ {u\over 2\sqrt {p}} +
\rd {\bar z}\
{\sqrt{\bar p}}\ \big( \sqrt{p {\bar p}}\ \re^{-2 \eta}-1\big)\,\Big]\ .
\eea
Here $r$  is given by  \eqref{sksksa}, and first term in the r.h.s. is the evaluated
form of the integral
${1\over 64}$\ $\int_{C_z}\rd z\times$ $ p^{-{5\over 2}}\,  (4pp''-5p'^2)$.

\subsection{Local IM in  quantum sine-Gordon model}

As was mentioned in  Introduction, the quantum sine-Gordon
model has infinitely many local integrals of motion,
\bea\label{ustsrslslssa}
{\mathbb I}_{2n-1}&=&2^{-4n}\
\int_0^R{\rd x\over 2\pi}\
\ \Big[\, \big(\partial_x\varphi-\partial_t\varphi\big)^{2n}+\ldots\,  \Big]\ ,\\
\label{ibar}{\bar {\mathbb I}}_{2n-1}&=& 2^{-4n}\ \int_0^R{\rd
x\over2\pi}\ \ \Big[\,
\big(\partial_x\varphi+\partial_t\varphi\big)^{2n}+\ldots\,
\Big]\ ,
\eea
where $\ldots$ stand for the terms involving higher
derivatives of $\varphi$, as well as the terms proportional to
powers of $\mu$. The displayed terms fix the normalization of
these operators. We will denote $I_{2n-1}=I_{2n-1}(k)$ and ${\bar
I}_{2n-1}={\bar I}_{2n-1}(k)$ the $k$-vacuum eigenvalues of the
operators \eqref{ustsrslslssa} and \eqref{ibar}, respectively.
In the CFT limit $M=0$ (i.e. at $\mu=0$ in \eqref{sg}) these
functions become polynomials in $k$ of the degree $n$\,\footnote{
In this limit \eqref{sg} acquires continuous symmetry with respect to any
shifts of the field $\varphi$, and the analytic continuation in
$k$ is no longer periodic.}, and the normalization in \eqref{ustsrslslssa}
is such that at $M=0$ we have
\bea\label{skklsklsa}
I_{2n-1}(k)\big|_{M=0}= {\bar I}_{2n-1}(k)\big|_{M=0}=
\big( {\textstyle {2\pi\over R}}\big)^{2n-1}\ \big[\,  (k\beta)^{2n}+\ldots\ \big]\ .
\eea

The expansions \eqref{ajahya}, \eqref{jsaskajsksa} are in agreement with the expected
asymptotic behavior of the $Q$-function of the 
quantum sine-Gordon model\ \cite{Bazhanov:1996dr,Lukyanov:2000jp}, with the
coefficients ${\mathfrak I}_{2n-1}$, ${\bar {\mathfrak I}}_{2n-1}$ and ${\mathfrak G}_{n}$,
${\bar {\mathfrak G}}_{n}$ being (up to normalization) the $k$-vacuum eigenvalues of
the local and non-local
\cite{Bernard:1990ys}
integrals of motion. In particular, we have
\bea\label{ssklsksaj}
{\mathfrak I}_{2n-1}
=C_n\
  I_{2n-1}\ ,\ \ \ \ \ \ \ \ \ \
{\bar {\mathfrak I}}_{2n-1}=
C_n\   {\bar I}_{2n-1}\ ,
\eea
where $C_n$ are constants, independent of $k$ and $R$. Their exact values can be found
by comparing the $s\to 0$ limit of ${\mathfrak I}_{2n-1}$ to \eqref{skklsklsa},
\bea\label{uysgksalk}
C_n={\Gamma\big(-
{2n-1\over 2\alpha}\big)\,
\Gamma\big(
{(2n-1)(\alpha+1)\over 2\alpha}\big)
\over 2\, \sqrt{\pi}\ n!}
\ \ \big(-{\textstyle{\alpha^2\over \alpha+1}}\big)^{n-1}\ \
\bigg[\,{m\over 8\sqrt{\pi}}\ \Gamma\big({\textstyle {\alpha+1\over 2\alpha}}\big)\,
\Gamma\big(-{\textstyle {1\over 2\alpha}}\big)\, \bigg]^{1-2 n}\ .
\eea
Here
\bea\label{kksksaoq}
m=2M\ \sin\big({\textstyle{\pi\over 2\alpha}}\big)\ .
\eea
At $\alpha>1$, this quantity coincides with the mass of the lowest soliton-antisoliton bound
state of the sine-Gordon model.

\section{ \label{Tfun} $T$-functions }

\subsection{Definition and relation with $Q_\pm$}

Let us consider the action of the symmetry transformation ${\hat \Omega}$,\ Eq.\eqref{kksksa}, on the
solution \eqref{kklssaklsa}.
As follows from\ \eqref{slslsla} and  analyticity
of ${\boldsymbol \Psi}_\pm$ and $Q_\pm$, the solution ${\boldsymbol \Xi}$ is entire function
of $\theta$ for any real $\rho$ and $\phi$. Therefore
analytic continuation of ${\boldsymbol \Xi}$ in $\theta$ can be used to specify another solution of the
linear problem\ \eqref{lp},
\bea\label{ystsls}
{\boldsymbol \Xi}_1(\rho,\phi\,|\,\theta)={\hat\Omega}[\,{\boldsymbol \Xi}\,](\rho,\phi\,|\,\theta)
\equiv
{\boldsymbol \Xi}\big(\rho,\phi+{\textstyle {\pi\over \alpha}}\,|\,\theta-
{\textstyle {\ri\pi\over \alpha}}\,\big)\ .
\eea
It is easy to see that for real $\theta$
and $|\phi|<{\textstyle\frac{\pi}{2(\alpha+1)}}$
\ \  $\ {\boldsymbol \Xi}_1$  grows at large $\rho$  as
\bea\label{uykklssaklsa}
{\boldsymbol \Xi}_1(\rho,\phi\,|\,\theta)\to  {\boldsymbol {\cal E}}_+(\rho,\phi\,|\,\theta)\ ,
\eea
where
\bea\label{qwaaszus}
{\boldsymbol {\cal E}}_+=-\ri\
\begin{pmatrix}
  \re^{-{\ri \alpha\phi\over 2}} \\
\re^{{\ri \alpha\phi\over 2}}
\end{pmatrix}
\ \exp\bigg[\, +{2\rho^{\alpha+1}\over \alpha+1}\
\cosh(\theta+\ri(\alpha+1)\phi\big)\, \bigg]\ .
\eea
Since
\bea\label{sokssa}
\det\big(\,{\boldsymbol \Xi},\, {\boldsymbol \Xi}_1\,\big)=
\det\big(\, {\boldsymbol {\cal E}}_-,\, {\boldsymbol {\cal E}}_+\,\big)=-2\,\ri \ ,
\eea
the pair  ${\boldsymbol \Xi}$ and ${\boldsymbol \Xi}_1$ forms a basis in the space of
solution of the linear problem\ \eqref{lp}.

Furthermore, applying the symmetry 
transformation ${\hat \Omega}^n$ with $n\in{\mathbb N}$, one can generate an infinite series
of  solutions
\bea\label{ystssklsasa}
{\hat\Omega}^n\big[\,{\boldsymbol \Xi}\,\big]\equiv
{\boldsymbol \Xi}\big(\rho,\phi+{\textstyle {\pi n\over \alpha}}\,|\,\theta-
{\textstyle {\ri\pi n\over \alpha}}\,\big)\ .
\eea
Of course, each of these solutions is a
linear combination of the basic solutions ${\boldsymbol \Xi}$ and  ${\boldsymbol \Xi}_1=
{\hat\Omega}[\,{\boldsymbol \Xi}\,]$. Using  Eq.\eqref{slslsla}, it is straightforward to show that
\bea\label{sksksaksa}
{\hat\Omega}^n\big[\,{\boldsymbol \Xi}\,\big]=
-T_{\frac{n-2}{2}}\big(\theta-
{\textstyle {\ri\pi(n+1)\over 2\alpha}}\,\big)
\ {\boldsymbol \Xi}(\rho,\phi\,|\,\theta)+
T_{\frac{n-1}{2}}\big(\theta-
{\textstyle {\ri\pi n\over 2\alpha}}\,\big)
\ {\boldsymbol \Xi}_1(\rho,\phi\,|\,\theta)
\ ,
\eea
where
\bea\label{usksksa}
&&T_j(\theta)={\ri \over 2\cos(\pi l)}\times\\
&&\ \ \  \Big[\,
Q_+\big(\theta+ {\textstyle {\ri\pi (2j+1)\over
2\alpha}}\big)\, Q_-\big(\theta- {\textstyle {\ri \pi(2j+1)\over 2\alpha}}\big)-
Q_+\big(\theta- {\textstyle {\ri\pi(2j+1)\over 2\alpha}}\big)\,Q_-\big(\theta+
{\textstyle {\ri \pi(2j+1)\over 2\alpha}}\big)\, \Big]\ .\nonumber
\eea
Note that
\bea\label{sksksksa}
T_{-{1\over 2}}(\theta)= 0\ ,\ \ \ \ T_{0}(\theta)= 1\ ,
\eea
where the second identity follows from the Quantum Wronskian relation\ \eqref{oosaopsa}.

$T_j(\theta)$ can be interpreted as Stokes  coefficients defining the
large $\rho$-behavior of the basic solution ${\boldsymbol \Xi}$.
Indeed, Eq.\eqref{sksksaksa} shows that for real $\theta$, and $\phi$ in the
domain
\bea\label{sslsal}
{\textstyle{\pi (2 n-3)\over 2(\alpha+1)}} <\phi<{\textstyle{ \pi (2 n-1)\over 2(\alpha+1)}}\ ,\ \ \ \
n=1,\, 2,\, \ldots
\eea
the asymptotic of   ${\boldsymbol \Xi}$ can be described as follows,
\bea\label{lsyat}
&&{\boldsymbol \Xi}(\rho,\,\phi\,|\,\theta)\to
(-1)^{[\frac{n-1}{2}]}\ T_{\frac{n-2}{2}}\big(\theta+
{\textstyle {\ri\pi (n-1)\over 2\alpha}}\,\big)
\ {\boldsymbol {\cal E}}_{-\sigma}(\rho,\phi\,|\,\theta)+\\ &&
\ \ \ \ \ \ (-1)^{[\frac{n}{2}]}\  T_{\frac{n-1}{2}}\big(\theta+
{\textstyle {\ri\pi n\over 2\alpha}}\big)\
\ {\boldsymbol {\cal E}}_{\sigma}(\rho,\phi\,|\,\theta)\ \ \ \ \ {\rm with}\ \ \ \
\sigma=(-1)^n\ \ \ \ \ {\rm as}\ \ \ \ \rho\to\infty\ .\nonumber
\eea
Here ${\boldsymbol {\cal E}}_\pm$ are given by Eqs.\eqref{qwaasz} and\ \eqref{qwaaszus}.
Of course, for   given  integer $n$  only one
term in \eqref{lsyat} is relevant whereas  another term should  be ignored.

\subsection{Fusion relations and $Y$-system}

The coefficients $T_j(\theta)$ in
\eqref{ystssklsasa} have useful interpretation in terms of the quantum sine-Gordon model.
$T_j(\theta)$ are the $k$-vacuum eigenvalues of the commuting ``transfer-matrices''
$\mathbb{T}_j (\theta)$,
the traces (over the $2j+1$ dimensional auxiliary spaces) of the quantum monodromy
matrices\ \cite{Baxter:1982,Sklyanin:1978}
(see also  Ref.\cite{Bazhanov:1996aq}, and Appendix B). 
The analytic properties of $T_j (\theta)$, which follow from
\eqref{usksksa}, are in agreement with expected analyticity of the $\mathbb{T}_j (\theta)$-operators in
sine-Gordon model \cite{Bazhanov:1996aq}.  The identities
\bea\label{skssa}
T_{1\over 2}(\theta)\ T_j\big(\theta+ {\textstyle {\ri \pi (2j+1) \over
2\alpha}}\big)=
T_{j-{1\over 2}}\big(\theta+ {\textstyle {\ri \pi (2j+2) \over
2\alpha}}\big)+
T_{j+{1\over 2}}\big(\theta+ {\textstyle {2\ri j \pi \over
2\alpha}}\big)\
\eea
(another simple consequence of \eqref{usksksa}) coincide with well known fusion relations
for the sine-Gordon transfer-matrices.

The fusion relations can be taken as the
starting point in purely algebraic derivation of the TBA equations.
In the sine-Gordon model with generic   $\beta^2$ the
TBA leads to an infinite  system of coupled integral equations,
the Takahashi-Suzuki system \cite{Takahashi:1972zza}, which
is somewhat difficult to deal with. However, at rational values of $\beta^2$ it truncates to
a finite system. Thus, at
\bea\label{alksjsakjs}
\alpha= N-1=2,\ 3,\, \ldots\ \ \ \ {\rm or}\ \ \ \ \ \
\alpha={\textstyle\frac{N}{2}}-1={\textstyle\frac{1}{2}},\
{\textstyle\frac{3}{2}},\ldots
\eea
Eq.\eqref{usksksa} dictates additional relation
\bea\label{sklaasuiasu}
T_{\frac{N}{2}}(\theta)=2\ \cos(2\pi k)+T_{\frac{N}{2}-1}(\theta)\ ,
\eea
which closes the fusion relations \eqref{skssa} within a finite number of functions $T_j(\theta)$,
$j=\frac{1}{2},\,1,\,\frac{3}{2},\,\ldots \frac{N}{2}-1$
(This truncation is discussed in Ref.\cite{Bazhanov:1996aq}). The standard way to derive the
associated finite system of TBA equations is to introduce the functions
\bea\label{jksjksajsa}
Y_j(\theta)&=&T_{j-\frac{1}{2}}(\theta)\, T_{j+\frac{1}{2}}(\theta)\, \ \ \ \ \ \ \ \ \ \ \ \
\big(\, j={\textstyle\frac{1}{2}},\, \ldots\,,\,{\textstyle\frac{ N}{2}}-1\,\big)\ ,\nonumber\\
Y_0(\theta)&=&0\ ,\\
{\bar Y}(\theta)&=&T_{\frac{N}{2}-1}(\theta)\ .\nonumber
\eea
As follows from the fusion relations and Eq.\eqref{sklaasuiasu}, $Y_j(\theta)$ satisfy
closed system of functional equations (the so-called $Y$-system)
\footnote{
Note that the TBA equations \eqref{isyststs}  for $l=0$, i.e.
$\re^{2\pi\ri k}=\ri$  in \eqref{isyststs},
correspond to the solution of MShG which remains finite at the apex of the  cone
${\mathbb C}_{\frac{\pi}{\alpha}}$ \eqref{cone}.
This case is of special
interest for the problem considered in \cite{Alday:2009yn,Alday:2009dv}.}:
\bea\label{isyststs}
&&Y_j\big(\theta+{\textstyle {\ri\pi\over 2\alpha}}\big)\,
Y_j\big(\theta-{\textstyle {\ri\pi\over 2\alpha}}\big)=
\big(\, 1+Y_{j-\frac{1}{2}}(\theta)\,\big)\,
\big(\,1+ Y_{j+\frac{1}{2}}(\theta)\,\big),\qquad
j={\textstyle\frac{1}{2}},\, \ldots\,,{\textstyle\frac{ N-3}{2}}\ ,\nonumber\\
&&Y_{\frac {N}{2}-1}\big(\theta+{\textstyle {\ri\pi\over 2\alpha}}\big)\,
Y_{\frac {N}{2}-1}\big(\theta-{\textstyle {\ri\pi\over 2\alpha}}\big)=
\big(\, 1+Y_{\frac {N-3}{2}}(\theta)\,\big)\,
\big(\,1+\re^{2\pi\ri k}\, {\bar Y}(\theta)\,\big) \big(\,1+
\re^{-2\pi\ri k}\,  {\bar Y}(\theta)\,\big)\ ,\nonumber\\
&&{\bar Y}\big(\theta+{\textstyle {\ri\pi\over 2\alpha}}
\big){\bar Y}\big(\theta-{\textstyle {\ri\pi\over 2\alpha}}\big)=
1+Y_{\frac {N}{2}-1}(\theta)\ .
\eea
This truncated $Y$-system coincides with the functional
form of the TBA equation of $D_N$ type \cite{Zamolodchikov:1991et} and
can be transformed to a set of the integral equations.
This  gives an alternative way to
reconstruct all the functions $T_j(\theta)$ and $Q_\pm(\theta)$ in the case
integer $2\alpha$\ \eqref{alksjsakjs}.

\subsection{Basic properties of $T_{\frac{1}{2}}$}

In what follows, the function $T(\theta,\,k)\equiv T_{1\over 2}(\theta)$ plays special role.
For future references, let us summarize here its general properties. All the statements
listed below are straightforward consequences of the definition the definition\ \eqref{usksksa} and
the properties of  $Q$-function.

\bigskip
\begin{itemize}

\item  For real $k$,  $T(\theta,\,k)$ is an entire function of $\theta$, even and periodic in this variable,
\bea\label{kslkska}
&&T(-\theta, k)=T(\theta,\, k)\, ,\nonumber\\
&&T\big(\theta+{\textstyle{\ri\pi (\alpha+1)\over \alpha}},\, k\, \big)=
T(\theta, k)\ .
\eea

\item  For real $k$,  $T(\theta,\,k)$ is a real analytic function of  $\theta$,
\bea\label{uyskslkska}
T^*(\theta,\,k\, )=T(\theta^*,\,k)\, \ .
\eea

\item It is even periodic function of $k$,
\bea\label{aksjsakj}
T(\theta,\,k)=T(\theta,-k)\ ,\ \ \ \ \ \ T(\theta,\, k+1)=T(\theta,\,k)\ .
\eea

\item  It satisfies the Baxter's  $T-Q$ equation
\bea\label{yrsksksa}
T(\theta,\, k)\ Q(\theta,\,  k)=
Q\big(\theta+{\textstyle{\ri\pi \over \alpha}},\,  k \big)
+Q\big(\theta-{\textstyle{\ri\pi \over \alpha}},\,  k \big)\ .
\eea
\item
For $\alpha>1$ and $\Re e (\theta)\to\pm \infty$
\bea\label{sssksa}
T(\theta,\,k )\to \exp\bigg[\,
{4\, {\hat r}\over \cos({\pi\over 2\alpha})}\, \cosh(\theta)
\,\bigg] \ \ \ \ \ \ {\rm for}\ \ \ \ \ \
\big|\,\Im m (\theta)\,\big|< {\textstyle{\pi (\alpha+1)\over 2\alpha}}\, .
\eea
\end{itemize}
Here and bellow we use the notation,
\bea\label{usysssksaswer}
{\hat r}=
-{ \pi^{3\over 2}\ \alpha 
\over (\alpha+1)\Gamma(-{1\over 2\alpha})\Gamma({\alpha+1\over 2\alpha})}\   s^{1+\alpha}
= \frac{m R}{8}\ ,
\eea
where $m$ is the same as in \eqref{kksksaoq}.

\section{\label{GLM}Inverse scattering problem for MShG}

\subsection{Gel'fand-Levitan-Marchenko equation}

As follows from
the   asymptotic formula   \eqref{kklssaklsa},
the composition of
symmetry transformations \eqref{kksksa} and   \eqref{gasfar}
${\hat \Pi}\circ {\hat\Omega}$,
acts irreducibly   on the solution ${\boldsymbol \Xi}$,
\bea\label{sskkjksasaq}
{\hat\Pi}\circ {\hat\Omega}\big[\,{\boldsymbol \Xi}\,\big]:=
{\boldsymbol \Xi}
\big(\rho, \phi+{\textstyle {\ri\pi\over \alpha}}\,\big|\, \theta-
 {\textstyle {\ri\pi (\alpha+1)\over \alpha}}\big)=
-\ri\ \sigma^3\ {\boldsymbol \Xi}(\rho, \phi\,|\,\theta)\ .
\eea
Combining this equation with \eqref{sksksaksa} one obtains
\bea\label{ayatksksaksa}
 &&(\,-\ri \sigma^3\,)^{n}\
{\boldsymbol \Xi}(\rho,\phi\,|\,\theta+\ri \pi n\,)=
-T_{\frac{n-2}{ 2}}\big(\theta-
{\textstyle {\ri\pi(n+1)\over 2\alpha}}\,\big)
\ {\boldsymbol \Xi}(\rho,\phi\,|\,\theta)-\\
&&\ \ \ \ \ \ \ri\
T_{\frac{n-1}{ 2}}\big(\theta-
{\textstyle {\ri\pi n\over 2\alpha}}\,\big)\ \sigma^3\
\ {\boldsymbol \Xi}(\rho,\phi\,|\,\theta+\ri\pi )
\ \ \ \ \ \ (n=1,\,2\ldots)\ .
\nonumber
\eea
For $n=2$,
with  $T_0 =1$ and the periodicity condition for $T_{1\over 2}$, Eq.\eqref{kslkska}, taken into account,
Eq.\eqref{ayatksksaksa} becomes a simple difference equation,
\bea\label{osislkslksklsa}
{\boldsymbol \Xi}(\rho,\phi\,|\,\theta+\ri\pi)-
{\boldsymbol \Xi}(\rho,\phi\,|\,\theta-\ri \pi )=\ri\
T_{\frac {1}{2}}(\theta)\ \sigma^3\
\ {\boldsymbol \Xi}(\rho,\phi\,|\,\theta )\ .
\eea

Our next goal is  to transform  \eqref{osislkslksklsa} into an
integral equation defining ${\boldsymbol \Xi}$.
Let us note that the coordinates $(\rho,\phi)$ on ${\mathbb C}_{\frac{\pi}{\alpha}}$, \eqref{cone},
appear in \eqref{osislkslksklsa} as parameters. For our analysis here, 
it will be convenient to use slightly
different coordinates on ${\mathbb C}_{\frac{\pi}{\alpha}}$. We redefine the angular coordinate $\phi$ by
shifting it by half the period,
\bea\label{asssusas}
\phi\to\phi+{\pi\over 2\alpha}\ ,
\eea
and use, instead of \ ${\cal M}_z^{(0)}$, the chart ${\cal M}_z^{(+)}$ defined by the same conditions as in
\ \eqref{skklslk} but in terms of the shifted angle $\phi$. 
Correspondingly, $z = \rho\,\re^{\ri\phi}$ will now
stand for the rotated complex coordinate,
\bea
z \to \re^{\frac{\ri\pi}{2\alpha}}\ z\,.
\eea
After this rotation we have $p(z) = -(z^{2\alpha}+s^{2\alpha})$. The minus sign here
does not affect the form of \ \eqref{shgz}, and in this and subsequent sections we set
\bea\label{sklsksksa}
p(z)=z^{2\alpha}+s^{2\alpha}\ .
\eea
In the rotated coordinate the zero of $p(z)$ appears
at the point $(\rho,\phi) = \big(s,\frac{\pi}{2\alpha}\big)\sim
\big(s,-\frac{\pi}{2\alpha}\big)$, as shown in Fig.2a. Actually, we will need 
the coordinate $w$ on ${\mathbb C}
_{\frac{\pi}{\alpha}}$
related to the new $z$ as in \eqref{jsaskajsksa} 
(it differs from $w$ in \eqref{uyslaskasa} essentially by a phase factor).
To
be precise, we set
\bea\label{systssaklslak}
w=w(z)={z^{\alpha+1}\over \alpha+1}+ \int_{z}^\infty\rd \zeta\ \Big[\,
\sqrt{\zeta^{2\alpha}+s^{2\alpha}}-\zeta^{\alpha}\,\Big]\ .
\eea
The image ${\cal M}_w^{(+)}$ of the chart ${\cal M}_z^{(+)}$ in the $w$-plane is shown in Fig.2b.
In ${\cal M}_w^{(+)}$ the apex of the cone has the coordinates $(w_0, w_0)$ with
\bea\label{ussyt}
w_0=w(0)=
-\frac{2\, {\hat r}}{\sin(\frac{\pi}{\alpha})}\ ,
\eea
while
\bea\label{usksaklssa}
w\big(s\, \re^{\pm \frac{\ri\pi}{2\alpha}}\big)=-{\hat r}\ \cot\big({\textstyle\frac{\pi}{2\alpha}}\big)
\pm\ri\  {\hat r}\ ,
\eea
where  notation\ \eqref{usysssksaswer} is used. In the remaining part of this paper we discuss in terms
of these redefined coordinates $z$ (or $w$), unless stated otherwise.
\begin{figure}[!ht]
\centering
\includegraphics[width=10  cm]{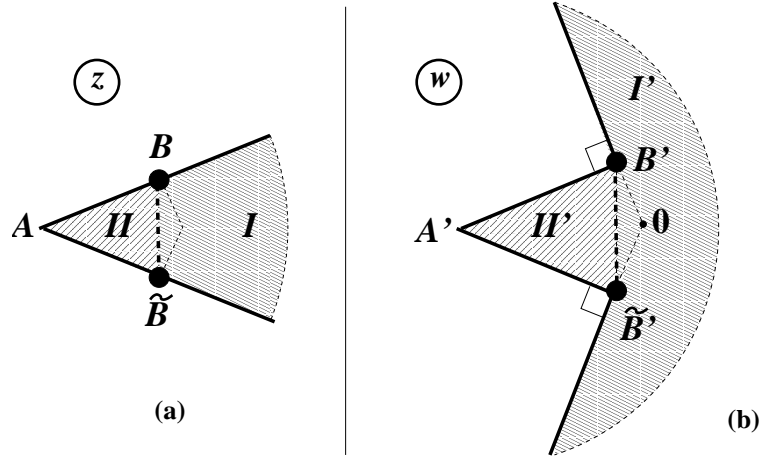}
\caption{(a) The chart $\mathcal{M}_{z}^{(+)}$. Two boundary rays (solid lines) are identified.
The dots indicate position of the zero of $p(z)$. (b) The image $\mathcal{M}_{w}^{(+)}$ of
$\mathcal{M}_{z}^{(+)}$ under the map \eqref{systssaklslak}. The segment $A'{\tilde B}'$ is identified
with $A'B'$, and the boundary line from ${\tilde B}'$ to infinity is identified with the line
from $B'$ to infinity. Specific regions in this chart are discussed in the text.}
\label{fig2}
\end{figure}

In the new variables $(w,{\bar w})$
the  solution ${\boldsymbol \Xi}(\rho,\phi\,|\,\theta )$ of the linear problem\ \eqref{lp}
can be written as
\bea\label{slssk}
{\boldsymbol \Xi}(\rho,\,\phi\,|\,\theta)=
\exp\Big[\, {\textstyle {\sigma^3\over 4}\ \big(\,
\ri\pi+\log\big({{\bar p}\over p}\big)\, \big)}\, \Big]\
{\hat {\boldsymbol \Xi}}\big(w,{\bar w}\,\,|\,\theta-
{\textstyle \frac{\ri\pi(\alpha+1)}{2\alpha}}\,\big)\ .
\eea
Here  ${\hat {\boldsymbol \Xi}}(w,{\bar w}\,|\,\theta)$ is the solution   of the linear problem
\bea\label{isusyssoiso}
&&
\Big( \partial_w +
{\textstyle{1\over 2}}\ \partial_w{\hat \eta}\ \sigma^3- \re^{\theta}\ \ \ \big[\,
\sigma^+\ \re^{{\hat \eta}}+ \sigma^-\re^{-{\hat \eta}}\, \big]\, \Big)\ {\hat {\boldsymbol \Xi}}=0\ ,
\\
&&\Big(\,
{ \partial}_{\bar w} -
{\textstyle{1\over 2}}\ {\partial}_{\bar w}{\hat \eta}\ \sigma^3-
 \re^{-\theta} \big[\, \sigma^-\, \re^{{\hat \eta}}+ \sigma^+\
\re^{-{\hat \eta}}\, \big]\, \Big)\ {\hat  {\boldsymbol \Xi}}=0\ , \nonumber
\eea
associated with  ShG equation\ \eqref{luuausay},
satisfying (for real $\theta$) the asymptotic condition,
\bea\label{sjksjssajkj}
{\hat  {\boldsymbol \Xi}}\to
\begin{pmatrix}
\ 1\\
-1
\end{pmatrix}\ \ \exp\big(\, - w\,\re^{\theta} -  {\bar  w}\, \re^{-\theta}\, \big) \ 
\ \ \ \ \ {\rm as}\ \ \
w\in {\cal M}_w^{(+)}\to\infty\ .
\eea
Note that ${\hat {\boldsymbol \Xi}}(w,{\bar w}\,|\,\theta)$ is nothing but   conventional Jost solution
for\ \eqref{isusyssoiso} (see, e.g., \cite{Faddeev:1987ph}  for details).
The  main advantage of the coordinates $(w,{\bar w})$ is relatively simple form of the large-$\theta$
asymptotic behavior  of ${\hat  {\boldsymbol \Xi}}$.
Simple analysis of the
linear problem \eqref{isusyssoiso} shows that
\bea\label{sslalsaiau}
{\hat {\boldsymbol \Xi}}(w,{\bar w}\,|\, \theta)\   \exp\big( \, w\, 
\re^{\theta} +{\bar  w}\, \re^{-\theta}\,
\big)
\to \re^{\omega_\pm}\ \re^{\pm \frac{\hat\eta\sigma^3}{2}}\
\begin{pmatrix}
\ 1\\
-1
\end{pmatrix}\
\ \ {\rm as}\ \ \  \theta\to \pm  \infty\, ,
\eea
where $\omega_\pm=\omega_\pm (w,\,{\bar w})$  are local solutions of
the Laplace equation $\partial_w\partial_{\bar w}\,\omega_\pm=0$ (in fact, $\omega_\pm$ are
piecewise constants, see Eq.\eqref{lyyastsa} below).
Combining the difference equation \eqref{osislkslksklsa}  with  the asymptotic behavior
\eqref{sslalsaiau} and the analyticity condition  for  ${\hat {\boldsymbol \Xi}}$,
it is straightforward to show that
\bea\label{ksjasssalas}
{\hat {\boldsymbol \Xi}}(w,{\bar w}\,|\, \theta)=
\begin{pmatrix}
\ \kappa_+ X_+(w,{\bar w}\,|\,\theta)\\
- \kappa_-  X_-(w,{\bar w}\,|\,\theta)
\end{pmatrix}\ \ \exp\big(\, w\, \re^{\theta} + {\bar  w}\, \re^{-\theta}\, \big)\ ,
\eea
where  $\kappa_\pm=\kappa_\pm (w, {\bar w})$ do not depend
on  $\theta$, while
$X_\pm $
are   solutions  of  linear integral  equations
\bea\label{tuyskksksa}
X_\pm (w,{\bar w}\,|\,\theta)=
1\pm \int_{-\infty}^{+\infty}{\rd\theta'\over 4\pi}\ \
\tanh\big({\textstyle \frac{\theta-\theta'}{2}}\big)\
D(w,{\bar w}\,|\,\theta')\ X_\pm (w,{\bar w}\,|\,\theta')
\eea
with the kernel
\bea\label{llksas}
D(w,{\bar w}\,|\,\theta)=
 T_{\frac{1}{2}}\big(\theta+{\textstyle{\ri\pi (\alpha+1)\over 2\alpha}}\big)
\ \exp\big( -2 w\, \re^{\theta}  -2{\bar  w}\, \re^{-\theta}\, \big)
\ .
\eea
With  Eqs.\eqref{sslalsaiau},\,\eqref{ksjasssalas} and
\bea\label{sslllllaqa}
\det\Big(\,{\hat {\boldsymbol \Xi}}(w,{\bar w}\,|\, \theta+{\textstyle\frac{\ri\pi}{2}}\big),\,
\sigma^3\ {\hat {\boldsymbol \Xi}}(w,{\bar w}\,|\, \theta-{\textstyle\frac{\ri\pi}{2}}\big)\,  \Big)=2
\eea
one obtains
\bea\label{kksklksa}
\kappa_\pm =d^{-\frac{1}{2}}\ \bigg[\,\frac{1-d_-^2}{1-d_+^2}\,\bigg]^{\pm \frac{1}{4}}\ ,
\eea
where
\bea\label{iusysksksksa}
d_\pm =\int_{-\infty}^{+\infty}{\rd\theta\over 4\pi}\ \ D(
w,{\bar w}\,|\,\theta)\
X_\pm (w,{\bar w}\,|\, \theta)
\eea
and
\bea\label{slaksas}
d\equiv {\textstyle\frac{1}{2}}\ \Big[\, X_+\big(\theta+{\textstyle\frac{\ri\pi}{2}}\big)
X_-\big(\theta-{\textstyle\frac{\ri\pi}{2}}\big)+
X_+\big(\theta-{\textstyle\frac{\ri\pi}{2}}\big)
X_-\big(\theta+{\textstyle\frac{\ri\pi}{2}}\big)\, \Big]=1-d_+d_-\ .
\eea
In these equations and below we omit the argument $(w,\wb)$ in $X_\pm$
and $d_\pm$, $\kappa_\pm$. Comparing the $\theta\to\pm\infty$ limits of \eqref{tuyskksksa} with
\eqref{sslalsaiau}, one can express the solution ${\hat \eta}$ of
ShG equation \eqref{luuausay} itself, as well as $\omega_\pm$, in terms of $d_\pm$,
\bea\label{lkslslsa}
\re^{2{\hat\eta}}=\frac{(1+d_+)(1+d_-)}{ (1-d_+)(1-d_-)}
\eea
and
\bea\label{lyyastsa}
\re^{\omega_\pm}=1\pm \frac{d_+-d_-}{d}\ .
\eea

Note that the conjugation condition
\bea\label{slakjsakjs}
{\hat {\boldsymbol \Xi}}^*(w,{\bar w}\,|\, \theta)=-\sigma^1\
{\hat {\boldsymbol \Xi}}(w,{\bar w}\,| -\theta^*)
\eea
implies the relations
\bea\label{sllauaassa}
X_-(w,{\bar w}\,|\,\theta)=X_+^*(w,{\bar w}\,|-\theta^*)\ ,\ \ \ \ \ \ \ \kappa_-=\kappa_+^*\ ,
\eea
and hence
\bea\label{jkajsakas}
d_-=d_+^*\, ,\ \ \  \ \ \ \ \ \ \ \ \ \ \
\omega_-=\omega_+^*\ .
\eea

Eq.\eqref{llksas} can be rewritten   in the form
\bea\label{llksasudys}
D(w,{\bar w}\,|\,\theta)=
T^{(\rm reg)}_{\frac{1}{2}}\big(\theta+{\textstyle{\ri\pi (\alpha+1)\over 2\alpha}}\big)\
\exp\big( -2 w_s\, \re^{\theta}  -2{\bar  w}_s\, \re^{-\theta}\, \big)
\ ,
\eea
where (see Eq.\eqref{usksaklssa})
\bea\label{sklslak}
w_s(z)=
\int^z_{s\re^{-\frac{\ri\pi}{2\alpha}}}\rd\zeta\ \sqrt{z^{2\alpha}+s^{2\alpha}}=
w(z)+{\hat r}\ \tan\big({\textstyle\frac{\pi}{2\alpha}}\big)-
\ri\ {\hat r}\ ,
\eea
and
\bea\label{klkslaksakla}
T^{(\rm reg)}_{\frac{1}{2}}(\theta)= T_{\frac{1}{2}}(\theta)\  \exp\bigg[\,
-{4\, {\hat r}\over \cos({\pi\over 2\alpha})}\, \cosh(\theta)
\,\bigg]\ .
\eea
Advantage of using the function $T^{(\rm reg)}(\theta)$ is that it
is well defined at any  $\alpha>0$, including the points
$\alpha^{-1}=1,\,3,\ldots$\,, as opposed to both $Q_\pm(\theta)$ and $T(\theta)$, which
at these points are
defined  only modulo overall factor
$\exp(\,\text{const}\, \theta)$.
Eq.\eqref{llksasudys} makes it explicit
that the kernel of   integral equations\ \eqref{tuyskksksa}
is well defined at  any  $\alpha>0$, including the integer points.
For example, in the case $\alpha=1$, it follows from
Eqs.\eqref{trsksksa},\,\eqref{sksklsskla},\,\eqref{yrsksksa} and
\eqref{klkslaksakla} that
\bea\label{llsssasla}
T^{(\rm reg)}_{\frac{1}{2}}(\theta)&=&\exp\bigg\{\, \int_{-\infty}^{\infty}
\frac{\rd t}{ 2\pi\cosh(\theta-t)}\times\\
&&
\log\Big[\,
\big(1+\re^{-\pi s^2\cosh(t)+2\pi\ri k}\, \big)\,
\big(1+\re^{-\pi s^2\cosh(t)-2\pi\ri k}\, \big)\,\Big]\, \bigg\}\ .\nonumber
\eea
For $k={1\over 4}$ this function, 
interpreted as a Stokes coefficient,  was found  in Ref.\cite{Gaiotto:2008cd}.
This result was also used in
Ref.\cite{Alday:2009yn}\footnote{In the notations of Ref.\cite{Gaiotto:2008cd,Alday:2009yn}:
$\gamma_1(\zeta)=T^{(\rm reg)}(\theta)$ 
with $\zeta\re^{-\ri\phi}=\re^{\theta}$,   $m=s^{2}\re^{\ri\phi}$.}.

Eqs.\eqref{llksasudys}-\eqref{klkslaksakla} and \eqref{sssksa} imply that
for
\bea\label{jaksjak}
\Re e\,(w)>-{\hat r}\ \tan\big({\textstyle\frac{\pi} {2\alpha}}\big)\ ,
\eea
and real $\theta$ the kernel $|D(w,{\bar w}\,|\theta)|$ is 
bounded by $\exp\big(-C\ \cosh(\theta)\big)$  with
positive constant $C$. In this domain of $w$ the
integral equations \eqref{tuyskksksa} have unique solutions obtainable by iterations.
With appropriate deformation  of the integration contours in \ \eqref{tuyskksksa} one can extend the domain
of applicability of the iterative solution  to the region  $I'\subset {\cal M}_w^{(+)}$ shown in    Fig.2b.
The most efficient way to solve  \eqref{tuyskksksa} at  the boundary of $I'$, i.e. on  the segment
$\Re e\,(w)=-{\hat r}\
\tan\big({\textstyle\frac{\pi} {2\alpha}}\big)$,
$\big|\Im m\,(w)\big|<{\hat r}$,
is  based on
the integral transformation in $\lambda=\re^\theta$
generated by the kernel  $\exp\big(\, \ri \xi\,  (\lambda-\lambda^{-1})\,\big)$.
The  transformation brings   \eqref{tuyskksksa}    into the form
of Gel'fand-Levitan-Marchenko equation (see e.g. \cite{Faddeev:1987ph} for details).
Important problem of  reconstruction of the Jost solution ${\hat {\boldsymbol \Xi}}(w,\,{\bar w}\,|\,\theta)$
in the whole domain ${\cal M}_w^{(+)}$ is beyond the scope of this paper.

\subsection{Solution of MShG equation}

Since ${\hat\eta}$ decays at $|w|\to\infty$, at large $|w|$ it
approaches certain solution ${\hat\eta}_1$ of the linearized equation 
$\partial_w\partial_{{\bar w}}\,{\hat\eta}_1 -4\,{\hat\eta}_1=0$. The form of this solution can be read
out from Eq.\eqref{lkslslsa},
\bea\label{sksl}
{\hat \eta}_1 =
\int_{-\infty}^\infty\frac{\rd \theta}{ 2 \pi}\
T_{\frac{1}{2}}
\big(\theta+{\textstyle{\ri\pi (\alpha+1)\over 2\alpha}}\big)\
\exp\big( -2 w\, \re^{\theta}  -2{\bar  w}\, \re^{-\theta}\, \big)\ .
\eea
Furthermore, iterations of \eqref{tuyskksksa} produce (through \eqref{iusysksksksa},
\eqref{slaksas}, and \eqref{lkslslsa}) systematic large-$|w|$ expansion of
${\hat\eta}$. In fact, one can guess the form of this expansion without explicit
calculations. As is known \cite{McCoy:1976cd},\cite{Zamolodchikov:1994uw}, the series
\bea\label{klklsaksa}
&&{\hat\eta} = \sum_{n=1}^{\infty}\,{\hat\eta}_{2n-1}\,,\\
&&{\hat\eta}_{2n-1} = {2\over 2n-1}\ \
\int_{-\infty}^\infty\prod_{j=1}^{2n-1}
\bigg[\, \frac{\rd\theta_j}{4\pi}\
\frac{{\cal T}(\theta_j)\,\exp\big( -2 w\, \re^{\theta}  -2{\bar  w}\, \re^{-\theta}\, \big)}
{\cosh\big({\theta_j-\theta_{j+1}\over 2}\big)}\, \bigg]\ ,\nonumber
\eea
with $\theta_{2n}\equiv\theta_1$, and
arbitrary function ${\cal T}(\theta)$, provides formal solution of the ShG equation \eqref{luuausay}.
Since the asymptotic form \eqref{lkslslsa}
fixes the solution uniquely, we conclude that the solution ${\hat\eta}$ we are interested
in is given (in certain domain of $w$ specified below) by the series \eqref{klklsaksa}
with
\bea
{\cal T}(\theta) = T_{\frac{1}{2}}
\big(\theta+{\textstyle{\ri\pi (\alpha+1)\over 2\alpha}}\big)\ .
\eea
This representation is very useful since at sufficiently large $w$ the series \eqref{klklsaksa}
converges fast.  In view of Eq.\eqref{sssksa},
the  multiple integrals in \eqref{klklsaksa}  converge only
when $w$ belongs to the  
domain \eqref{jaksjak}. With deformation  of the integration contours, the convergence
domain can be extended to
$I'\subset {\cal M}_w^{(+)}$ shown in Fig.\ref{fig2}b.
Thus, the  solution of
MShG\ \eqref{shgz}  in the domain $I\subset {\cal M}_z^{(+)}$ (Fig.\ref{fig2}a)  
can be written in the form
$\eta={\hat \eta}+\frac{1}{4}\ \log(p {\bar p})$, where
${\hat\eta}$ and $p$ are given by \eqref{klklsaksa} and
\eqref{sklsksksa}, respectively. It would be interesting to find similarly explicit expression
for ${\hat\eta}$ in the remaining part of ${\cal M}_z^{(+)}$.

As $|w|\to\infty$, the integral  \eqref{sksl} can be evaluated   by the saddle-point method.
As the result one  derives the  large-$\rho$  asymptotic of the   solution of MShG
\eqref{shgz},\,\eqref{kskssls},
\bea\label{oislsls}
\eta(\rho,\phi)&\to& {\textstyle \frac{1}{4}}\ \log \big(\, s^{4\alpha}-
2\,(s\rho)^{2\alpha}\,\cos(2\alpha\phi)+\rho^{4\alpha}\,\big)\\
&+&
T_{\frac{1}{2}}\big(\,\ri (\alpha+1)\phi\, \big)\
 {\re^{- \tau}\over \sqrt{2\pi \tau}}\ \Big(\,1+O(\tau^{-1})\,\Big)\
\ \ {\rm as}\   \ \tau={\textstyle\frac {4\rho^{\alpha+1}}{\alpha+1}}\to+\infty\ .\nonumber
\eea
In writing this equation we use the original polar
coordinates $(\rho,\,\phi)$ on the chart ${\cal M}_z^{(0)}$\ \eqref{skklslk},
 and  assume for simplicity that $\alpha>1$.
Eq.\eqref{oislsls} shows  that the $T$-function
 $T_{\frac{1}{2}}(\theta)$ determines the angular dependence of the sub-leading large-$\rho$ asymptotic
of the MShG solution. At $s=0$ and finite $\theta$ the function $T_\frac{1}{2}(\theta)$ becomes a constant,
\bea\label{kaskssa}
T_{\frac{1}{2}}(\theta)\big|_{s=0}=2\ \cos\big({\textstyle \frac{\pi (2l+1)}{2(\alpha+1)}}\big)\ ,
\eea
and
hence \eqref{oislsls}  generalizes  the well known  asymptotic formula  for the
Painlev\'e III transcendent\ \cite{McCoy:1976cd}.

\section{\label{MShSh}
MShG with $\alpha<-1$ and quantum sinh-Gordon model}

So far we were concentrating attention on MShG with $\alpha >0$.
In that case we had a freedom to adjust the asymptotic behavior
of $\eta$ at $z\to 0$ as in \eqref{kakskasosi}, with  the free
parameter $l$. If $\alpha < -1$, the situation is different:
asymptotic form of a regular solution is fixed at both
$z\to\infty$ and $z\to 0$. In fact, the roles of $z=0$ and $z=\infty$
can be interchanged by certain ``duality'' transformation. Conformal transformation
\begin{eqnarray}\label{duz}
{\tilde z} = \left[-\,\frac{s^\alpha\,z}{\alpha+1}\,\right]^{\alpha+1}
\end{eqnarray}
(with suitable shift of the field $\eta$) brings the MShG equation
to the original form, but with the parameters $\alpha, s$ replaced
by the ``dual'' values ${\tilde\alpha},\ {\tilde s}$ related to
the original ones as
\bea\label{dualpha}
(\alpha+1)\,({\tilde\alpha}+1)=1\,, \qquad \qquad\qquad  {\tilde s}^{{\tilde\alpha}+1} =
-\frac{s^{\alpha+1}}{\alpha+1}\ .
\eea
Note that in terms of the parameters $b$ and $\mu$, Eq.\eqref{alphab},
these relations faithfully reproduce the $b \to b^{-1}$ duality symmetry
of the quantum sinh-Gordon model \cite{Zamolodchikiv2001}. Below we argue that such solution
(more precisely, certain connection coefficient for solutions of
the linear problem \eqref{lp}) is related to the $Q$-function of
the sinh-Gordon model \eqref{shg}.

In this discussion we will use the chart ${\cal M}^{(+)}_z$ on
${\mathbb C}_{\frac{\pi}{\nu}}$, in which $p(z)$ has the form
\eqref{sklsksksa}, and associated polar coordinates $z =
\rho\exp({i\phi})$. We also use the notation
\bea\label{skklskj}
\nu=-{\alpha}\ .
\eea
We assume that MShG equation has solution
$\eta(\rho,\phi)$ which is real and continuous on ${\mathbb
C}_{\frac{\pi}{\nu}}$ except for the apex $\rho=0$, and has the
following properties (compare to the properties ${\bf i})-{\bf iv})$ in Section\,\ref{LInP})

${\bf i}')$
\begin{eqnarray}
\eta\big(\rho, \phi+{\textstyle\frac{\pi}{\nu}}\big) =
\eta(\rho,\phi)\,.
\end{eqnarray}

${\bf ii}')$
$\eta(\rho,\phi)$ are real-valued and finite
everywhere on the cone $\mathbb{C}_\frac{\pi}{\alpha}$, except
for the apex $\rho=0$.

\bigskip

${\bf iii}')$

\bea\label{kasosiisy} \eta(\rho,\phi )\to 0 \ \ \ \ \ \ \ \ {\rm
as}\ \ \ \ \rho\to \infty\ . \eea

${\bf iv}')$
\bea\label{kakosi} \eta(\rho,\phi)\to -\nu\  \log
(\rho)+O(1)\ \ \ \ \ \ \ \ {\rm as}\ \ \ \ \rho\to 0\ . \eea

\bigskip

\noindent The relevant linear   problem\ \eqref{lp} now is somewhat
simpler then that previously discussed. At $\nu>1$  we
introduce    two solutions of the linear problem, defined by the
asymptotic conditions at $\rho\to\infty$ and $\rho\to 0$,
\bea\label{uytkssksatr}
{{\boldsymbol
\Xi}}_+(\rho,\,\phi\,|\,\theta)\to
\begin{pmatrix}
1\\
-1
\end{pmatrix}\
\exp\Big[-2s^{-\nu}\ \rho\ \cosh(\theta+\ri\phi)\,\Big]\ \ \ \ \
\ {\rm as}\ \ \ \rho\to +\infty\ , \eea and \bea\label{skssksatr}
{ {\boldsymbol  \Xi}}_-(\rho,\,\phi\,|\,\theta)\to
\begin{pmatrix}
 \re^{-{\ri \nu \phi\over 2}} \\
\re^{{\ri \nu\phi\over 2}}
\end{pmatrix}
\ \exp\bigg[\, -{2\rho^{1-\nu}\over \nu-1}\
\cosh(\theta-\ri(\nu-1)\phi\big)\, \bigg]\ \ \ \ \ \ {\rm as}\ \
\ \rho\to 0\ . \eea We then define \bea\label{skklsa} {\cal
Q}(\theta)={\textstyle{1\over 2}}\
  \det\big({ {\boldsymbol \Xi}}_+,\, {  {\boldsymbol \Xi}}_-\big)\  .
\eea

By arguments parallel to the analysis in
\cite{Zamolodchikiv2001, Fateev:2005kx}, it is possible to
establish the following properties of this connection coefficient:

\begin{itemize}

\item
 ${\cal Q}(\theta)$ is  an entire  function of  $\theta$ with the symmetries
\bea\label{skskskksaksal}
{\cal Q}(\theta)={\cal Q}(-\theta)\ ,\
\ \  \ \ \ \ {\cal Q}^*(\theta)= {\cal Q}(\theta^*)\ .
\eea

\item  ${\cal Q}(\theta)$  satisfies
the Quantum Wronskian relation:
\bea\label{mnmslksjs}
{\cal
Q}\big(\theta- {\textstyle{{\rm i}\pi\over 2}}\big) {\cal
Q}\big(\theta+{\textstyle{{\rm i}\pi\over 2}}\big) - {\cal
Q}\big(\theta+{\textstyle{{\rm i}\pi (\nu-2)\over 2\nu}}
\big){\cal Q}\big(\theta-{\textstyle{{\rm i}\pi (\nu-2) \over
2\nu}}\big) =1\ .
\eea

\item  As the function of the complex
$\theta$, ${\cal Q }(\theta)$  is free of zeroes in the strip
$\big|\Im m\,\theta\big|<{\pi\over 2}+\epsilon$ for some finite
$\epsilon>0$.

\item For
$\big|\Im m\,\theta\big|<{\pi\over 2}+\epsilon$ and $\Re e\,
\theta\to+\infty$
\bea\label{ytsalksj}
{\cal Q }(\theta)=
\exp\bigg[- { {2\ {\hat r}\over \sin({\pi\over \nu})}}\
\re^\theta+O( \re^{-\theta})\,\bigg]\ ,
\eea
where\footnote{Note the
similarity in the definitions of  ${\hat r}$ for  $\nu=-\alpha>1$
and for $\nu=-\alpha<0$  \eqref{usysssksaswer}. It   is defined
to be positive for any  $|\alpha|>1$.}
\bea\label{usysssksas}
{\hat r}={ \pi^{3\over 2}\ \nu\over (\nu-1)\Gamma({1\over
2\nu})\Gamma({\nu-1\over 2\nu})}\   s^{1-\nu}\ .
\eea

\end{itemize}

\noindent Let us introduce the function $\varepsilon(\theta)$ through the
relation \bea\label{lsks}
{\cal Q}\big(\theta+ {\textstyle{{\rm
i}\pi\over 2}}\big)\, {\cal Q }\big(\theta- {\textstyle{{\rm
i}\pi\over 2}}\big)=1+\re^{-\varepsilon(\theta)}\ .
\eea
With the
analytic properties listed above, it is straightforward to
transform the difference equation\ \eqref{mnmslksjs}\ into 
integral equation for $\varepsilon(\theta)$ (see
Ref.\cite{Zamolodchikov:2000kt}),
\bea\label{iikksksa}
\varepsilon(\theta)-8\,{\hat r}\ \cosh(\theta)+
\int_{-\infty}^\infty{\rd\theta'\over 2\pi}\ \Phi(\theta-\theta')\
\log\big(1+\re^{-\varepsilon(\theta')}\,\big)
\eea
with the
kernel
\bea\label{iusyslslsa}
\Phi(\theta)={4\,\sin({\pi\over\nu})\ \cosh(\theta)\over
\cosh(2\theta)-\cos({2\pi\over\nu})}\ .
\eea
Then
\bea\label{isuysyussksklsa} \log{\cal Q}(\theta)=- {4\, {\hat
r}\over \sin({\pi\over \nu})}\ \cosh(\theta)+
\int_{-\infty}^\infty{\rd\theta'\over 2\pi}\ {
\log\big(1+\re^{-\varepsilon(\theta')}\big)\over
\cosh(\theta-\theta')}\ .
\eea

As was argued in Ref.\cite{Lukyanov:2000jp},
Eq.\eqref{isuysyussksklsa} gives the   $Q$-function for  the
quantum sinh-Gordon model in the finite-size geometry
\eqref{pbc}, provided ${\hat r}=\frac{mR}{8}$, with $m$
interpreted as the mass of the sinh-Gordon particle, and $\nu$
related to the sinh-Gordon coupling constant $b$ as in
\eqref{alphab}, i.e.
\bea\label{skklskjiusy}
\nu= b^{-2}+1\ .
\eea

\begin{figure}[!ht]
\centering
\includegraphics[width=12  cm]{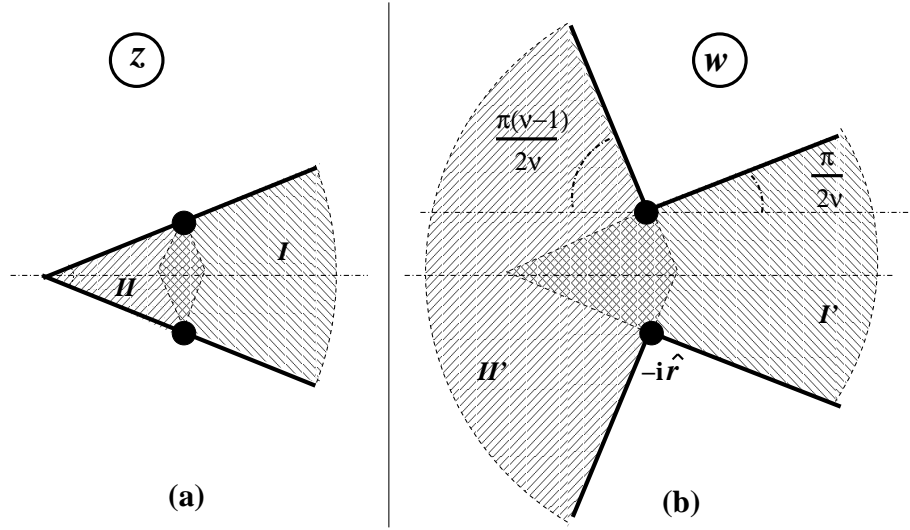}
\caption{The chart $\mathcal{M}_{z}^{(+)}$ (a), and its image $\mathcal{M}_{w}^{(+)}$ under the
map \eqref{skskskjla} (b) in the case $\nu = -\alpha >0$. }
\label{fig3}
\end{figure}

It is instructive to review the above statements in terms of the coordinate
\bea\label{klsklskjsla}
w=w(z)=\int\rd z\  \sqrt{z^{-2\nu}+s^{-2\nu}}
\eea
which brings the MShG equation to the form of the conventional ShG equation \eqref{luuausay}
for ${\hat\eta} = \eta-\frac{1}{4}\,\log(p{\bar p})$.
We fix  the integration constant  in \eqref{klsklskjsla} in such a way that
\bea\label{skssasal}
w\big(\, s\,\re^{\pm \ri{\pi\over 2\nu}}\,\big)= \pm\ri\,{\hat r}\ .
\eea
Explicitly,
\bea\label{skskskjla}
w(z)=-{\hat r}\
\cot\big({\textstyle{\pi\over 2\nu}}\big)+
z\, s^{-\nu}
\  {}_2F_1\big(-{\textstyle{1\over 2}},\,-{\textstyle{1\over 2\nu}},\,
1 - {\textstyle{1\over 2\nu}},
-\big({\textstyle{z\over s}}\big)^{-2 \nu}\ \big)\ .
\eea
The chart ${\cal M}^{(+)}_z$ and  its  $w$-image ${\cal M}^{(+)}_w$
are shown in Fig.\ref{fig3}a and Fig.\ref{fig3}b.
The asymptotic conditions ${\bf iii}')$ and ${\bf iv}')$ above simply mean that ${\hat\eta}$
decays at $|w| \to \infty$ in both regions $I'$ and  $II'$ in Fig.3b.
Eq.\eqref{skklsa} can be equivalently written as
\bea\label{skklsaossuu}
{\cal Q}(\theta)={{1\over 2}}\
\exp\bigg[-
{ {4\, {\hat r}\over \sin({\pi\over \nu})}}\
\cosh(\theta)\, \bigg]\ \  \det\big(\,{\hat {\boldsymbol \Xi}}_+,\, {\hat  {\boldsymbol \Xi}}_-\big)\  ,
\eea
where
${\hat {\boldsymbol \Xi}}_\pm$  are conventional  Jost solutions\ \cite{Faddeev:1987ph} for
the linear problem \eqref{isusyssoiso} satisfying the asymptotic conditions
\bea\label{uytkssksa}
{\hat {\boldsymbol \Xi}}_+(w,{\bar w}\,|\,\theta)\to
\exp\big(-w\,\re^\theta -{\bar  w}\,\re^{-\theta}\,\big)\
\begin{pmatrix}
1\\
-1
\end{pmatrix}\ \ \ \ \ \ {\rm as}\ \ \ \Re e(w)\to +\infty\ ,
\eea
and
\bea\label{skssksa}
{\hat {\boldsymbol  \Xi}}_-(w,{\bar w}\,|\,\theta)\to
\exp\big(\, w\,\re^\theta + {\bar w}\,\re^{-\theta}\,\big)\
\begin{pmatrix}
1\\
1
\end{pmatrix}\ \ \ \ \ \ {\rm as}\ \ \ \Re e(w)\to -\infty\ .
\eea

In this picture the duality \eqref{duz} is quite evident. The rotation
$w\to -w$ interchanges domains $I'$ and $II'$ in Fig.\ref{fig3}b , which is equivalent
to the change of the parameters
\bea\label{oyreyskksksa}
\nu^{-1}\to 1-\nu^{-1}\ , \qquad {\hat r}\to{\hat r}
\eea
identical to \eqref{dualpha} (Note that \eqref{usysssksas} is invariant under the transformation
\eqref{dualpha}).
It is easy to check that under this duality
\bea\label{skssaaap}
{\hat {\boldsymbol \Xi}}_\pm (w,{\bar w}\,|\,\theta) \leftrightarrow
\sigma^3\ {\hat {\boldsymbol  \Xi}}_\mp (-w, -{\bar w}\,|\,\theta)\ ,
\eea
and hence the function ${\cal Q}(\theta)$ does not change when its parameters are transformed
as in \eqref{oyreyskksksa}. Thus, it is invariant with respect to the $b \to b^{-1}$ duality, as
the sinh-Gordon $Q$-function should be.

The asymptotic expansions
at $\theta\to+\infty$ and $\theta\to-\infty$ of ${\cal Q}(\theta)$ generate the
vacuum eigenvalues of the local  integral of motions
in the quantum sinh-Gordon model. Namely \cite{Lukyanov:200
0jp},
\bea\label{skasklaslqt}
\log{\cal Q}\sim-C_0\ \re^\theta-\sum_{n=1}^\infty C_n\, I_{2n-1}\ \re^{-(2n-1)\theta}\ \ \ \
{\rm as}\ \ \ \ \ \theta\to+\infty\ ,\\
\log{\cal Q}\sim-C_0\ \re^{-\theta}-\sum_{n=1}^\infty C_n\, {\bar I}_{2n-1}\ \re^{(2n-1)\theta}\ \ \ \
{\rm as}\ \ \ \ \ \theta\to-\infty\ ,\nonumber
\eea
where $I_{2n-1}$ and ${\bar I}_{2n-1}$ are the  vacuum eigenvalues of  the local  integral of motions
${\mathbb I}_{2n-1}$ and ${\bar {\mathbb I}}_{2n-1}$\ $(n=1,\,2,\,\ldots)$ normalized as in\ \eqref{ustsrslslssa}.
The constant $C_n$ are given by \eqref{uysgksalk} with $\alpha=-\nu$, where $m$ is interpreted as the
mass of the sinh-Gordon particle.
From Eq.\eqref{isuysyussksklsa} we have
\bea\label{kklsksajl}
C_n\ I_{2n-1}&=&C_0\ \delta_{n,1}+(-1)^n\ \int_{-\infty}^\infty{\rd\theta\over \pi}\
\re^{(2n-1)\theta}\ \log\Big(1+\re^{-\varepsilon(\theta)}\,\Big) \ ,\\
C_n\ {\bar I}_{2n-1}&=&C_0\ \delta_{n,1}+(-1)^n\ \int_{-\infty}^\infty{\rd\theta\over \pi}\
\re^{-(2n-1)\theta}\ \log\Big(1+\re^{-\varepsilon(\theta)}\,\Big)\ .\nonumber
\eea
On the other hand, straightforward WKB analysis allows one to express \eqref{kklsksajl} in terms of the classical conserved
charges for MShG equation,
\bea\label{hjsjahsaa}
C_n\ I_{2n-1}&=&{(-1)^n
\over 2n-1}\
\int_{-\infty}^\infty \Big[\,\rd w\ {\hat P}_{2n}+
\rd {\bar w}\ {\hat  R}_{2n-2}\,\Big] \ ,\\
C_n\ {\bar I}_{2n-1}&=&{
(-1)^n\over 2n-1}\  \int_{-\infty}^\infty\Big[\, \rd {\bar w}\ {\hat {\bar P}}_{2n}+
\rd { w}\ {\hat {\bar  R}}_{2n-2}\, \Big]\ .\nonumber
\eea

Further analysis is similar to the one presented in Section\,\ref{GLM} for   $\alpha>0$.
The novel  property is existence of two Jost solutions.
For this reason one has to introduce two $T$-functions ${\cal T}_{\pm}$,
\bea\label{usysslsl}
{\cal T}_{+}(\theta)\ {\cal Q}(\theta)&=&{\cal Q}\big(\theta+
{\textstyle {\ri\pi\over\nu}}\big)+{\cal Q}\big(\theta-
{\textstyle {\ri\pi\over\nu}}\big)\ ,\\
{\cal T}_{-}(\theta)\ {\cal Q}(\theta)&=&{\cal Q}\big(\theta+{\textstyle {\ri (\nu-1)\pi\over\nu}}\big)+
{\cal Q}\big(\theta-
{\textstyle {\ri\pi(\nu-1)\over\nu}}\big)\ .\nonumber
\eea
Note that the duality transformation \eqref{oyreyskksksa} interchanges the $T$-functions,
\bea\label{isyslslsasa}
{\cal T}_{\pm}(\theta) \leftrightarrow {\cal T}_{\mp}(\theta)\ .
\eea
The   ShG solution ${\hat \eta}$  still can be  written
as the series \eqref{klklsaksa}, but the choice of the function ${\cal T}(\theta)$ is different
for the two parts $I'$ and  $II'$ of the chart ${\cal M}^{(+)}_w$ in Fig.3b. Thus, for $w \in I'$ we
have
\bea\label{klklsaksaaaa}
{\hat\eta} = \sum_{n=1}^{\infty}\,\,
{2\over 2n-1}\ \
\int_{-\infty}^\infty\prod_{j=1}^{2n-1}
\bigg[\, \frac{\rd\theta_j}{4\pi}\
\frac{{\cal T}_{+}(\theta_j)\,\exp\big( -2 w_+\, \re^{\theta}  -2{\bar  w_+}\, \re^{-\theta}\, \big)}
{\cosh\big({\theta_j-\theta_{j+1}\over 2}\big)}\, \bigg]\ ,
\eea
where
\bea\label{ssksllksa}
w_+=
{\hat r}\ \cot\big({\textstyle{\pi\over 2\nu}}\big)+w\ .
\eea
For $w \in II'$ one has to replace ${\cal T}_+(\theta)$ by ${\cal T}_{-}(\theta)$, and
$w_+$ by
\bea
w_-= -{\hat r}\
\tan\big({\textstyle{\pi\over 2\nu}}\big)-w\ .
\eea
Since the union $I' \cup II'$ covers  the whole chart ${\cal M}^{(+)}_w$,
combination of these two representations provide the solution $\eta$ on
the whole cone ${\mathbb C}_{\frac{\pi}{\nu}}$.

\section{\label{disss} Discussion}

In this paper we have described  relation between the
classical MShG equation and its linear problem, on one hand, and
quantum sine- and sinh-Gordon models on the other. This relation
generalizes the relation \cite{Dorey:1998pt,Bazhanov:1998wj} between ordinary differential
equations \cite{Voros:1992,Voros:1999bz} and integrable structures of Conformal Field Theories
\cite{Bazhanov:1994ft,Bazhanov:1996dr,Bazhanov:1998dq} to the massive case. We believe it also brings useful
insight into the emergence of TBA equations in recent analysis of
the MShG equation \cite{Gaiotto:2008cd, Gaiotto:2009hg, Alday:2009yn, Alday:2009dv, Alday:2010vh}.

The discussion in this paper is in terms of the $Q$- and
$T$-functions of the quantum models \eqref{sg} and \eqref{shg},
which are defined through the Bethe Ansatz for the vacuum states.
More generally, these functions have to be understood as the
vacuum eigenvalues one-parameter families of commuting operators
$\mathbb{Q}(\theta)$ and $\mathbb{T}(\theta)$. In integrable
lattice models (e.g. XXZ and XYZ chains) these operators were
discovered in pioneering works of Baxter \cite{Baxter:1971, Baxter:1982}. When an
integrable quantum field theory emerges as continuous (scaling)
limit of such lattice systems, it inherits these operators.
However, for many reasons (including subtleties of the continuous
limit) it is desirable to have constructions of these operators
directly in field theoretic terms. There was some progress in
this direction. Thus, in Ref.\cite{Bazhanov:1994ft} the operators $\mathbb{T}_j
(\theta),\ j = \frac{1}{2},\, 1,\,\frac{3}{2},\,\ldots$ were constructed
explicitly for massless (conformal) field theory, as traces of
quantum monodromy matrices over $2j+1$ dimensional auxiliary
spaces. This construction of $\mathbb{T}_j(\theta)$ admits more
or less direct extension to the massive sine-Gordon model and its
reductions \cite{Bazhanov:1996aq}. In both
massless and massive cases it allows one to establish directly basic properties
of these operators, which we summarize in Appendix B. In particular,
one can argue that all $\mathbb{T}_j(\theta)$ are entire functions of
$\theta$, in
the sense that all their simultaneous eigenvalues are entire
functions.
Furthermore, in the massless case similar construction (with
auxiliary space supporting representation of $q$-oscillator
algebra) exists for the $Q$-operator \cite{Bazhanov:1996dr,Bazhanov:1998dq}. It is
plausible that it also admits generalization to the massive
case, but the details were never elaborated. However, both the massless construction
and the lattice theory suggest certain properties of the sine-Gordon $Q$-operator.
Thus, one expects that $\mathbb{Q}(\theta)$ is entire function of $\theta$ as well.
This operator commutes with all $\mathbb{T}_j(\theta')$, and satisfies
the famous $T-Q$ equation of Baxter,
\bea\label{usytstlsalssa}
{\mathbb T}_{\frac{1}{2}}(\theta)\
{\mathbb Q}(\theta)= {\mathbb Q}\big(\theta+{\textstyle
\frac{\ri\pi }{\alpha}} \big)+ {\mathbb Q}\big(\theta+{\textstyle
\frac{\ri\pi }{\alpha}} \big)\ ,
\eea
familiar from the lattice theory \cite{Baxter:1971, Baxter:1982}.
This equation is finite difference analog
of a second order differential equation. Since in the sine-Gordon model
$\mathbb{T}(\theta)$ is periodic function of $\theta$ (see Eq.\eqref{skksklsklsa}),
one expects to have two ``Bloch wave''  solutions $\mathbb{Q}_\pm (\theta)$,
\bea\label{ilsalssaisus}
&&{\mathbb Q}_{+}\big(\theta+{\textstyle \frac{\ri\pi (\alpha+1) }{\alpha}} \big)=
{\mathbb U}\ {\mathbb Q}_{+}(\theta)\ , \\
&&{\mathbb Q}_{-}\big(\theta+{\textstyle \frac{\ri\pi (\alpha+1) }{\alpha}} \big)=
{\mathbb U}^{-1}\ {\mathbb Q}_{-}(\theta)\ ,
\nonumber
\eea
with some unitary operator $\mathbb{U}$ commuting with $\mathbb{Q}_\pm(\theta)$.
Again, by comparison to the massless limit \cite{Bazhanov:1996dr,Bazhanov:1998dq,Bazhanov:1996aq},
it is natural to identify $\mathbb{U}$
with the Flouquet-Bloch operator associated with the discrete symmetry of the sine-Gordon
theory,
\bea\label{kssusuausyt}
{\mathbb  U}\,\varphi(x,t)\,  {\mathbb  U}^{-1}=\varphi(x,t)+2\pi/\beta\ .
\eea
Since $\mathbb{T}_\frac{1}{2}(\theta)$ is invariant with respect to
the charge conjugation \eqref{ystlsslsllsa}, the operators $\mathbb{Q}_{+}
(\theta)$ and $\mathbb{Q}_{-}(\theta)$ are related to each other by this
symmetry transformation, Eq.\eqref{isuysystulsalssa} (this symmetry was already taken
into account in writing \eqref{ilsalssaisus}), so one can deal with one
independent operator, say $\mathbb{Q} \equiv \mathbb{Q}_+$. Additional piece of
analytic information -- the asymptotic behavior
\bea\label{qasss}
\log\mathbb{Q}(\theta) \sim \text{const}\ \re^{\pm\theta} + O(1)
\eea
as $\Re e(\theta) \to\pm \infty$ in the strips $H_\pm$\ \eqref{oskskskjsa} 
-- can be inferred from the massless
$Q$-operators \cite{Bazhanov:1996dr,Bazhanov:1998dq}. 
These assumptions and  some of their simple consequences (in particular,
precise form of \eqref{qasss}) are summarized in  Appendix B.

Once the sine-Gordon $Q$-function $Q(\theta,k)$ is understood as the
$k$-vacuum eigenvalue of the $Q$-operator, the question arises about
its eigenvalues associated with the excited states. Basic properties of such
eigenvalues can be inferred from \eqref{trelsalssa}-\eqref{usystjsajksajka}.
Can the excited-state eigenvalues be also related to integrable classical
equations? In the massless case it is known how to modify  the
Schr${\rm \ddot o}$dinger equation \eqref{schroedinger} to accommodate for the excited states
\cite{Bazhanov:2003ni}. In the massive theory this is  interesting
open question. Of course, generalization of the DDV equation \eqref{ksalaks} to the excited
states is well known \cite{Destri:1997yz,Feverati:1998dt}.

\section*{Acknowledgments}

The authors are grateful to  Volodya Bazhanov,   Alyosha Litvinov and Fedya Smirnov for discussions.
Our special thanks  to Greg Moore, who encouraged us
to look for interpretation  of the results of Refs.\cite{Gaiotto:2008cd,Gaiotto:2009hg}
in terms of 2D  quantum  integrability.

\bigskip

\noindent This  research was supported in part by DOE grant
$\#$DE-FG02-96 ER 40949.

\bigskip

\noindent Research of ABZ falls within the framework of the Federal Program
``Scientific and Scientific-Pedagogical Personnel of Innovational Russia''
on 2009-2013 (state contract No. 02.740.11.5165) and supported  by RFBR
initiative interdisciplinary project grant 09-02-12446-OFI-m.

\section*{Appendix}

\appendix

\section{Derivation of Eqs.\eqref{slslsaiaua}-\eqref{sklsklsalas}}

Here we sketch derivation of the properties
\eqref{slslsaiaua}-\eqref{sklsklsalas} of $Q_\pm$.

\noindent
Eqs.\eqref{sklsaksklsa} follow from\ \eqref{lslsak},\,
\eqref{kkslsssal} and easily established relation
\bea\label{skskskla}
\sigma^1\,
{\boldsymbol \Xi}^*(\rho,\phi\,|-\theta)=
-{\boldsymbol \Xi}(\rho,\phi\,|\,\theta)\qquad \text{at}
\quad
\Im m(\theta)=0\ .
\eea

To prove the  quasiperiodicity   \eqref{slslsaiaua}   we use the
evident relation
\bea\label{kssk}
Q_-(\theta)= W \ \det\Big[\,
{\boldsymbol \Xi}(\rho,\phi\,|\,\theta),\,{\boldsymbol \Psi}_-(\rho,\phi\,|\,\theta)\, \Big]\ ,
\eea
where $W=-\cos(\pi l)$.
Then
\bea\label{salslsa}
&&Q_-\big(\theta-{\textstyle {\ri
\pi (\alpha+1)\over 2\alpha}}\big)=
W\times\\
&& \det\Big[\, {\boldsymbol \Xi}\big(\rho,\phi+{\textstyle {\pi\over 2\alpha}}\,|\,
\theta
-{\textstyle {\ri \pi (\alpha+1)\over
2\alpha}}\,\big)\, ,\,{\boldsymbol \Psi}_-\big(\rho,\phi+{\textstyle {\pi\over
2\alpha}}\,|\, \theta
-{\textstyle
{\ri \pi (\alpha+1)\over 2\alpha}}\,\big)\, \Big]\ .\nonumber \eea
The following
formula is immediate  consequence of
Eqs.\eqref{ahaytatissu},\,\eqref{sssalksa},
\bea\label{slalalsk}
{\boldsymbol \Psi}_-
\big(\rho,\phi+{\textstyle{\pi\over\alpha}}\,|\,\theta-{\textstyle{\ri\pi(\alpha+1)\over\alpha}}\,\big)=
\re^{\ri\pi l}\  \sigma^3\ {\boldsymbol \Psi}_-(\rho ,\phi\, |\,\theta)\ .
\eea
It
is also straightforward to prove similar relation for the
solution ${\boldsymbol \Xi}$,
\bea\label{lakksa}
{\boldsymbol \Xi}\big(\rho,\phi+{\textstyle
{\pi\over \alpha}}\,|\, \theta-{\textstyle {\ri \pi(\alpha+1)\over
\alpha}}\,\big)= -\ri\, \sigma^3\ {\boldsymbol \Xi}(\rho,\phi\,|\, \theta)\ .
\eea
Using
\eqref{slalalsk} and \eqref{lakksa} one has \bea\label{sjsksalk}
&&Q_-\big(\theta
-{\textstyle {\ri
\pi(\alpha+1)\over 2\alpha}}\,\big)=
-\ri\, \re^{\ri\pi l}\ W\times\\
&& \det\Big[\,
\sigma^3\,
{\boldsymbol \Xi}\big(\rho,\phi+{\textstyle {\pi\over 2\alpha}}\,
|\, \theta
-{\textstyle {\ri \pi (\alpha+1)\over 2\alpha}}\, \big)\,
,\, \sigma^3\, {\boldsymbol \Psi}_-\big(\rho,\phi-{\textstyle {\pi\over 2\alpha}}\,|\,
\theta+{\textstyle {\ri \pi (\alpha+1)\over 2\alpha}}\, \big)\, \Big]\nonumber\\
&&\ \ \ \ \ \ \ \ \ \  \ \ \ \ \ \ \ \ \ \ \ \  =\ri\,\re^{\ri\pi l}\ Q_-\big(\theta
+{\textstyle {\ri (\alpha+1) \pi\over 2\alpha}}\, \big)\ . \nonumber \eea

For given values of $\rho$ and $\phi$,  the solution
${\boldsymbol \Xi}$ \eqref{kklssaklsa}, considered
as the function of complex $\theta$, is
analytic in the strip
$-{\textstyle{\pi\over 2}}
- (\alpha+1)\,\phi\leq \Im m(\theta)\leq {\textstyle{\pi\over 2}}-(\alpha+1)\, \phi$.
Since $ {\boldsymbol \Psi}_-$ is entire function of $\theta$, one concludes
that
$Q_+(\theta)$ is analytic in the strip
$|\Im m (\theta)|<\pi+{\pi\over 2\alpha}$, and hence due to the
 quasiperiodicity   \eqref{slslsaiaua} it is entire function of $\theta$.
Since $Q_\pm$ and ${\boldsymbol \Psi}_\pm$ are entire functions of $\theta$,
\  ${\boldsymbol \Xi}$\ \eqref{slslsla} is
an entire function as well.

To prove the Quantum Wronskian relation\ \eqref{oosaopsa}  we use
\eqref{slslsla} to obtain
\bea\label{sskssa}
{\boldsymbol \Xi}(\rho,\phi\mp
{\textstyle {\pi\over 2\alpha}}\, |\, \theta\pm {\textstyle
{\ri\pi\over 2\alpha}}) &=& Q_-(\theta\pm  {\textstyle
{\ri\pi\over 2\alpha}})\
{\boldsymbol \Psi}_+(\rho,\phi\mp
{\textstyle {\pi\over 2\alpha}}\,|\,\theta\pm {\textstyle {\ri\pi\over 2\alpha}})\nonumber\\
&+& Q_+(\theta\pm {\textstyle {\pi\over 2\alpha}})\
{\boldsymbol \Psi}_-(\rho,\phi\mp {\textstyle {\pi\over 2\alpha}}\,|\,\theta\pm
{\textstyle {\ri\pi\over 2\alpha}})\ .
\eea
From\
\eqref{ahaytatissu} we have
 \bea\label{xsskssa} {\boldsymbol \Xi}\big(\rho,\phi\mp
{\textstyle {\pi\over 2\alpha}}\, |\, \theta\pm {\textstyle
{\ri\pi\over 2\alpha}}\big) =\Big[\, Q_-(\theta\pm {\textstyle
{\ri\pi\over 2\alpha}})\ {\boldsymbol \Psi}_+(\rho,\phi\,|\,\theta)+ Q_+(\theta\pm
{\textstyle {\pi\over 2\alpha}})\ {\boldsymbol \Psi}_-(\rho,\phi\, |\, \theta)\,
\Big]\ ,
\eea
and hence
\bea\label{ksksaksa}
&&\det\Big[\,
{\boldsymbol \Xi}(\rho,\phi-{\textstyle {\pi\over 2\alpha}}\,|\,\theta+ {\textstyle
{\ri \pi\over 2\alpha}}),\, {\boldsymbol \Xi}(\rho,\phi+{\textstyle {\pi\over
2\alpha}}\,|\,\theta- {\textstyle {\ri \pi\over 2\alpha}})\,
\Big]=\det\big[\, {\boldsymbol \Psi}_+, {\boldsymbol \Psi}_-\,\big]
\times\nonumber\\
&& \big(\, Q_-(\theta+ {\textstyle {\ri\pi\over
2\alpha}})\, Q_+(\theta- {\textstyle {\ri \pi\over 2\alpha}})-
Q_-(\theta- {\textstyle {\ri\pi\over 2\alpha}})\, Q_+(\theta+
{\textstyle {\ri \pi\over 2\alpha}})\, \big)\ . \eea
Using
normalization condition
\eqref{slslsa} and easily established
relation
\bea\label{skssakl}
\det\Big[\,
{\boldsymbol \Xi}(\rho,\phi-{\textstyle {\pi\over 2\alpha}}\,|\,\theta+ {\textstyle
{\ri \pi\over 2\alpha}}),\, {\boldsymbol \Xi}(\rho,\phi+{\textstyle {\pi\over
2\alpha}}\,|\,\theta- {\textstyle {\ri \pi\over 2\alpha}})\,
\Big]=-2\ri\ , \eea
one arrives at  \eqref{oosaopsa}.

Finally, Eqs.\eqref{sklsklsalas} are immediate consequences of\ \eqref{mahah}.

\section{$T$ and $Q$ -operators in sine-Gordon model}

\subsection{$T$-operators}

Let ${\mathbb C}$ and ${\mathbb P}$ be unitary  operators of charge
conjugation and parity transformation in the sine-Gordon model \eqref{sg},
\bea\label{ystlsslsllsa}
{\mathbb C}\,\varphi(x,t)\, {\mathbb C}&=&-\varphi(x,t)\ ,\\
\label{ystlsslsllsaa}
{\mathbb P}\,\varphi(x,t)\, {\mathbb P}&=&\varphi(-x,t)\ .
\eea

Integrability of the quantum sine-Gordon model can be expressed in terms of
family of operators (``transfer-matrices'') ${\mathbb T}_j(\theta)\,, \ \theta\in \mathbb{C}\,,
\ j=\frac{1}{2},\,1,\, \frac{3}{2},\,\ldots$, having the following properties ($\alpha=\beta^{-2}-1)$:

\begin{itemize}
\item  Mutual commutativity
\bea\label{lsalssa}
[\,{\mathbb T}_j(\theta),\, {\mathbb T}_{j'}(\theta')\,]=0\ .
\eea

\item  ${\mathbb  U}$ and ${\mathbb  C}$ invariance
\bea\label{uislsalssa}
[\,{\mathbb T}_j(\theta)\,,\, {\mathbb  U}\,]=[\,{\mathbb T}_j(\theta)\,,\, {\mathbb  C}\,]=0\ .
\eea

\item  Parity transformation
\bea\label{ulsalssa}
{\mathbb  P}\, {\mathbb T}_j(\theta)\,  {\mathbb  P}={\mathbb T}_j(-\theta)\ .
\eea

\item Hermiticity
\bea\label{ksksaksla}
{\mathbb T}^\dagger_j(\theta)= {\mathbb T}_j(\theta^*)\ .
\eea

\item Periodicity
\bea\label{skksklsklsa}
{\mathbb T}_j\big(\theta+{\textstyle \frac{\ri\pi (\alpha+1)}{\alpha}} \big)=
{\mathbb T}_j(\theta)\ .
\eea

\item Fusion relation
\bea\label{skssaissu}
{\mathbb T}_{\frac{1}{2}}(\theta)\ {\mathbb T}_j\big(\theta+ {\textstyle {\ri \pi (2j+1) \over
2\alpha}}\big)=
{\mathbb T}_{j-\frac{1}{2}}\big(\theta+ {\textstyle {\ri \pi (2j+2) \over
2\alpha}}\big)+
{\mathbb T}_{j+\frac{1}{2}}\big(\theta+ {\textstyle {2\ri j \pi \over
2\alpha}}\big)\ .
\eea

\item  Asymptotic at real $\theta$:
\bea\label{llsslsa}
\log {\mathbb T}_{\frac{1}{2}}(\theta)\sim \sum^\infty_{n=0}\
2\,(-1)^n\, \sin\big({\textstyle\frac{\pi (2n-1)}{2\alpha}}\big)\ C_n\times
\begin{cases}
& {\mathbb I}_{2n-1}\ \re^{-(2n-1)\theta} \ \ \ \ {\rm as}\ \ \ \ \theta\to+\infty\\
& {\bar{\mathbb I}}_{2n-1}\ \ \re^{(2n-1)\theta}\ \ \ \ \  {\rm as}\ \ \ \ \theta\to-\infty
\end{cases}\  .
\eea
Here ${\mathbb I}_{-1}={\bar {\mathbb I}}_{-1}=
\frac{R}{2\pi}$ is a $c$-number, while  ${\mathbb I}_{2n-1} \
(n=1,\,2\ldots)$ are local IM\ \eqref{ustsrslslssa},
and the constants $C_n$ are given by\ \eqref{uysgksalk}.

\end{itemize}

\noindent
The operators $\mathbb{T}_j (\theta)$ can be thought of as the continuous limits
of the Baxter's commuting transfer-matrices \cite{Baxter:1982}, or traces of monodromy matrices
of the quantum inverse scattering method \cite{Sklyanin:1978}. Construction of these operators directly
in continuous quantum field theory is outlined in
Refs.\cite{Bazhanov:1994ft,Bazhanov:1996dr, Bazhanov:1998dq, Bazhanov:1996aq}.

\subsection{$Q$-operators}

Here we summarize expected properties of 
the operators ${\mathbb Q}_{\pm}(\theta)$. Eqs.\eqref{trelsalssa},\,\eqref{yislsalssa}
just recapitulate what was already suggested in Section\,\ref{disss}. The rest follows from the 
$T-Q$ equation
\eqref{usytstlsalssa}, the defining relations \eqref{ilsalssaisus}, and the 
asymptotic\ \eqref{qasss}.

\begin{itemize}
\item  Commutativity
\bea\label{trelsalssa}
[\,  {\mathbb Q}_\pm(\theta),\, {\mathbb T}_{j}(\theta')\,]=
[\,  {\mathbb Q}_\pm(\theta),\,{\mathbb Q}_\pm(\theta')\,]=
[\,  {\mathbb Q}_+(\theta),\,  {\mathbb Q}_-(\theta')\, ]=0\ .
\eea
\item
${\mathbb  U}$ invariance
\bea\label{yislsalssa}
[\,{\mathbb Q}_\pm(\theta)\,,\, {\mathbb  U}\,]=0\ .
\eea

\item Quantum Wronskian relation
\bea\label{ksalkaisiu}
{\mathbb Q}_+\big(\theta+{\textstyle \frac{\ri\pi }{2\alpha}} \big)\
{\mathbb Q}_-\big(\theta-{\textstyle \frac{\ri\pi }{2\alpha}}\big)\,
-
{\mathbb Q}_+\big(\theta-{\textstyle \frac{\ri\pi }{2\alpha}} \big)\
{\mathbb Q}_-\big(\theta+{\textstyle \frac{\ri\pi }{2\alpha}}\big)
=
{{\mathbb U}}^{-1}-{\mathbb U}\ ,
\eea
where $\mathbb{U}$ is the field translation \eqref{kssusuausyt}.

\item Charge and parity conjugations
\bea\label{isuysystulsalssa}
{\mathbb  C}\,{\mathbb Q}_\pm (\theta)\, {\mathbb  C}=
{\mathbb Q}_\mp (\theta)\ , \qquad
{\mathbb  P}\,{\mathbb Q}_\pm (\theta)\, {\mathbb  P}=
{\mathbb Q}_\mp(-\theta)\ .
\eea

\item Hermiticity
\bea\label{ystrksksaksla}
{\mathbb Q}^\dagger_\pm(\theta)= {\mathbb Q}_\pm(\theta^*)\ .
\eea

\item Leading asymptotic  in the strips $H_\pm$\eqref{amaayay} (we assume $\alpha\not =1,\,3,\,5\ldots$)
\bea\label{usystjsajksajka}
{\mathbb Q}_+(\theta) &\to&{{\mathbb U} }^{\pm\frac{1}{2}}\ \  {{\mathbb S} }^{\frac{1}{2}}
\ \ \ \, \exp\bigg[\,{MR \, \re^{\theta\mp {\ri\pi(1+\alpha)\over
2\alpha}}\over 4\cos({\pi\over 2\alpha})}\,\bigg] \quad \text{as}\quad \Re e\,(\theta)\to +\infty\,,\\
{\mathbb Q}_+(\theta)&\to& {{\mathbb U} }^{\pm\frac{1}{2}}\
{{\mathbb S} }^{-\frac{1}{2}}\ \ \exp\bigg[\,{MR \,
\re^{-\theta\pm {\ri\pi(1+\alpha)\over 2\alpha}}\over
4\cos({\pi\over 2\alpha})}\,\bigg]\quad \text{as}\quad \Re e\,(\theta)\to -\infty\ .\nonumber
\eea
Here  ${{\mathbb S}}$ is some operator which is invariant with respect to the
$\mathbb{P}$ and $\mathbb{U}$ symmetries,
\bea\label{kkssaasiuat}
[\,{\mathbb S},\, {\mathbb P}\,]=[\,{\mathbb S},\, {\mathbb U}\,]=0\ ,
\eea
and satisfies the relations
\bea\label{aisulsksjasuu}
{\mathbb S}^{-1}={\mathbb C}\, 
{\mathbb S}\, {\mathbb C}\ ,\ \ \ \ \ \ \ {\mathbb S}^{\dagger}={\mathbb S}\ .
\eea

\end{itemize}

At the moment we do not know physical interpretation of the operator $\mathbb{S}$, but regard it as very
interesting open question. In the massless case it is similar
to the Liouville
``reflection S-matrix'' \cite{Zamolodchikov:1995aa}\footnote{We note in this connection
that in the massless case the full spectrum of $\mathbb{S}$
can be extracted from recent remarkable paper\ \cite{Jimbo:2009ja}.}.

\end{document}